# Subsequent Dust-Fading Events of the Wolf-Rayet Star WR 59: A Continuous Visual Record

Rod Stubbings

*Tetoora Road Observatory, 2643 Warragul-Korumburra Road, Tetoora Road, VIC 3821, Australia*

*stubbo@dcsi.net.au*

**Abstract**

This paper presents long-term visual and space-based monitoring of the WC9d Wolf-Rayet star WR 59 from 2017 to 2026, documenting a profound physical transition from a stable baseline to an ongoing phase of continuous dust production. Visual observations using a 22-inch telescope initially established a four-year quiescent baseline at magnitude 13.9. Following the historic May 2021 discovery fade to magnitude 15.2, long-term nightly visual monitoring documented an accelerating sequence of twelve distinct fading episodes over subsequent years, characterised by progressively shorter inter-fade recovery intervals. Due to severe pixel crowding and flux contamination from a close 11.7 *V* companion star 11 arcseconds away, automated large-pixel survey pipelines remain completely blind to these high-amplitude fading events. Direct comparisons with NASA's *TESS* satellite photometry during quiet phases show that targeted visual magnitude estimates measure systemic baseline micro-variations with professional accuracy while bridging space mission coverage gaps. High-density observation blocks during the minimum floor reveal a highly dynamic, rapid, minute-by-minute bimodal oscillation and sub-minute flickering between magnitudes 15.0 and 15.2. At the time of writing, the system remains locked in an unprecedented deep minimum, varying between magnitudes 15.0 and 15.2 for a sustained period exceeding 102 consecutive days. Intensive visual monitoring is ongoing to record the eventual egress recovery profile.

**Keywords:** stars: individual (WR 59) – stars: Wolf-Rayet – techniques: visual – dust, extinction

## 1. Introduction

The discovery of the first-ever recorded dust-fading event of the Wolf-Rayet star WR 59 in May 2021 (Stubbings 2026) showed that this system can undergo dramatic visual declines. Historically, the star was documented as a stable baseline object maintaining a magnitude near 13.9 since its early cataloguing (van der Hucht et al. 1981), with its formal classification as a persistent dust maker (WC9d) established in subsequent catalogues (van der Hucht 2001). Optical absorption features in the spectrum suggest that WR 59 is a binary system (Williams et al. 2005). WR 59 sits 11 arcseconds away from a magnitude 11.7 *V* companion star, causing severe pixel crowding and flux contamination in automated and CCD data streams. However, continuous nightly monitoring from the Tetoora Road Observatory using high-magnification visual astronomy has revealed that the 2021 fade was not an isolated anomaly. Instead, it marked the beginning of a highly dynamic, active phase for the system.

To date, my ongoing visual observing programme has documented an additional twelve distinct fading events. As of November 2026, WR 59 is locked in an unprecedented deep minimum that has held variations between magnitudes 15.0 and 15.2 for more than 102 consecutive days (part of the 2026 Fade N complex). This paper presents the complete, unblended visual records from April 2017 to November 2026 of these subsequent fading events alongside high-density intra-night monitoring. These observations document how WR 59 has transitioned into a series of frequent, chaotic dust-making episodes that remain completely hidden from CCD data and automated large-pixel survey pipelines.

## 2. Visual Observations and Milestone Timeline

All observations in this campaign were conducted from the Tetoora Road Observatory using a 22-inch (0.56-m) telescope at 270× magnification. To map the long-term behaviour of WR 59, I maintained a continuous nightly visual observing programme alongside CCD *V*-band and space-based datasets. Figure 1 shows the calibrated light curve compiled by Sebastian Otero of the AAVSO International Variable Star Index (VSX; Watson et al. 2006) spanning from 2017 (JD 2457800) to early 2026 (JD 2461100). In this plot, my visual observations (SRX) establish the zero-point baseline. To account for contamination from a close magnitude 11.7 *V* companion star, mathematical offsets were applied to the other datasets, as shown in the plot legend. This includes space-based data from NASA's Transiting Exoplanet Survey Satellite (*TESS*), where the light curve documents the baseline state in Sector 11, the historic 2021 discovery fade in Sector 38, and subsequent quiescent baseline states in Sectors 65 and 100. Because my later 2026 visual data were not in the AAVSO database, the subsequent Fade M, Sector 102 coverage, and the current long Fade N sequence are excluded from this specific overarching plot. They are instead detailed individually in later sections.

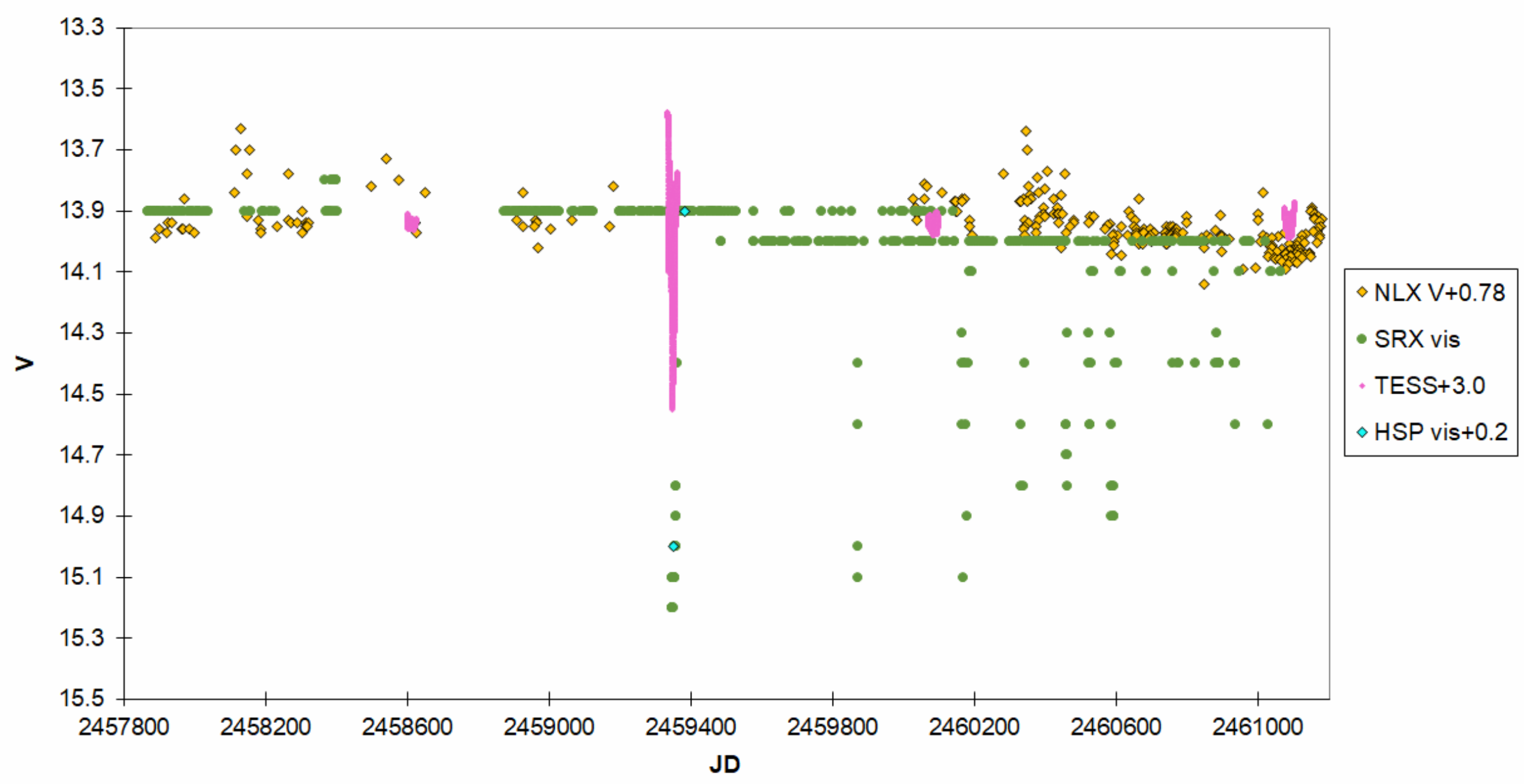

**Figure 1.** Calibrated multi-year historical light curve of the WC9d Wolf-Rayet star WR 59 spanning 2017 to early 2026 (JD 2457800 to JD 2461100). Unblended visual observations (SRX vis; green circles) establish the true, isolated physical baseline of the primary star near magnitude 13.9. To correct for severe spatial pixel crowding and flux contamination from the close magnitude 11.7 *V* companion star, mathematical offsets were applied to the comparison datasets as indicated in the legend. These include CCD *V*-band data records (NLX; yellow diamonds, offset by +0.78 mag), visual points (HSP vis; cyan diamonds, offset by +0.2 mag) and high-cadence space-based photometry from NASA's Transiting Exoplanet Survey Satellite (*TESS*; magenta lines, offset by +3.0 mag). The space-based data successfully frame the initial quiet baseline in Sector 11, the historic May 2021 discovery fade in Sector 38, and subsequent quiescent intervals in Sectors 65 and 100.

To contrast this calibrated view with raw archival records, Figure 2 shows the AAVSO light curve plotted over the exact same multi-year interval, from JD 2457800 to JD 2461100. In raw terms, the uncorrected CCD *V*-band data sit at a relatively flat baseline near magnitude 13.1. This artificial brightness stems from severe pixel crowding and flux contamination caused by the close magnitude 11.7 *V* companion star and WR 59. Physically isolating the star at the eyepiece reveals the true unblended visual baseline at magnitude 13.9, exposing the complete blindness of uncorrected CCD data to these deep fading events.

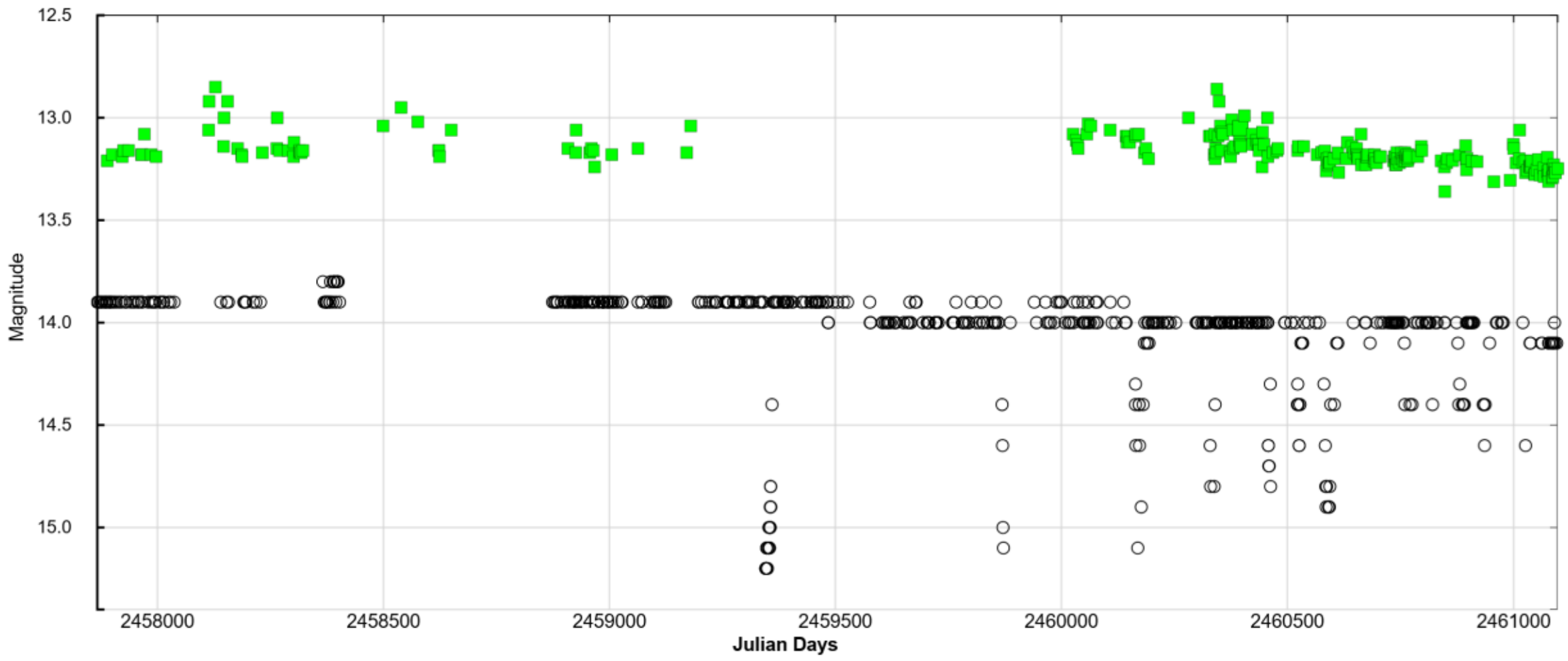

**Figure 2.** Comparison of uncorrected CCD records and visual estimates for WR 59 across the identical multi-year window from JD 2457800 to JD 2461100. Raw CCD *V*-band data (green squares) show an artificially bright baseline near magnitude 13.1 that remains completely unaffected by the fading events. This total blindness stems from severe pixel crowding and flux contamination from the close magnitude 11.7 *V* companion star. In stark contrast, physically isolating WR 59 at the eyepiece of a 22-inch telescope (black circles) reveals the true, unblended visual baseline at magnitude 13.9. This high-magnification optical isolation completely bypasses the companion contamination, revealing the true depth, duration, and structure of the deep fading events extending down to magnitude 15.2.

## 3. The Multi-Fade Chronology

Before evaluating individual fading events, it is vital to establish the system's long-term historical baseline. After regular monitoring began in April 2017 on JD 2457867, WR 59 showed exceptional stability for an interval of four years (1,480 days) leading up to its first active deep fade minimum. Across this extended window, which includes a notable field gap, the star remained relatively stable at its unblended visual magnitude of 13.9.

Crucially, the space-based light curve data right in the middle of this visual gap, specifically NASA's *TESS* Sector 11 spanning JD 2458596 to 2458623, verifies that the system remained entirely at its quiescent baseline. This coverage is critical because it rules out the possibility that any short-interval fades or dipping events occurred unobserved while the field was not monitored visually. This confirms that the frequent fading events seen later in the campaign were absent during these first four years of prolonged stability.

When analysing the comprehensive master log, a highly significant and physical transition stands out. The data revealed that the intervals between subsequent dust-fading events were not random. During the early years of activity, the star took its time, producing independent events separated by extended intervals. The long-term record documents this structural compression clearly in Table 1 with the interval from the minimum of Fade A to the minimum of Fade B spanning 525 days, which then shortened to 297 days between Fade B and Fade C, before shrinking to just 161 days between Fade C and Fade D. As the campaign advanced into the mid-2020s, the time between events compressed rapidly down to 63 days for Fade F and a compact 44 days for Fade K.

Intriguingly, just before the modern active state, the system showed a temporary structural relaxation as the intervals for Fades L and M widened slightly to 90 and 94 days, respectively. Rather than signalling a return to the historic quiescent state, this localised lengthening represents a final disruption of the star's independent fading behaviour, serving as the immediate physical prelude to the continuous, ongoing complex dust-production phase. To map how the system behaved once it broke out of this prolonged quiescent phase after the historic May 2021 discovery event, I compiled a master summary timeline showing the intervals between every independent fade right up to the start of the current active epoch. Listing some minimum steps as double entries helps document the exact time of the minimum and the duration of these obscured phases for each fading event.

**Table 1.** Master chronological summary of independent fading events (Fades A to M)

| Observed Event | Julian Date (JD) | Visual Mag | Interval (Days) |
|---|---|---|---|
| Campaign Initiation | 2457867.0100 | 13.9 | Baseline |
| Pre-Gap Boundary | 2458402.9040 | 13.9 | 535.8940 |
| [Unobserved Field Gap] | - | - | - |
| Visual Recovery Interval | 2458874.1400 | 13.9 | 471.2360 |
| Deep Fade A Minimum | 2459346.9010 | 15.2 | 472.7610 |
| Deep Fade B Minimum | 2459871.9170 | 15.1 | 525.0160 |
| Deep Fade C Minimum | 2460169.0640 | 15.1 | 297.1470 |
| Fade D Ingress Floor | 2460330.1920 | 14.8 | 161.1280 |
| Fade D Egress Verification | 2460338.1260 | 14.8 | 7.9340 |
| Fade E Wave Minimum | 2460462.8770 | 14.8 | 124.7510 |
| Fade F Core Dip | 2460526.0510 | 14.6 | 63.1740 |
| Fade G Ingress Floor | 2460585.9110 | 14.9 | 59.8600 |
| Fade G Egress Verification | 2460592.9140 | 14.9 | 7.0030 |
| Fade H Ingress Floor | 2460760.0410 | 14.4 | 167.1270 |
| Fade H Egress Verification | 2460775.8970 | 14.4 | 15.8560 |
| Fade I Wave Minimum | 2460820.9010 | 14.4 | 45.0040 |
| Fade J Ingress Floor | 2460879.0250 | 14.4 | 58.1240 |
| Fade J Egress Verification | 2460891.0430 | 14.4 | 12.0180 |
| Fade K Wave Minimum | 2460935.9060 | 14.6 | 44.8630 |
| Fade L Wave Minimum | 2461026.1790 | 14.6 | 90.2730 |
| Deep Fade M Ingress Floor | 2461121.0403 | 15.0 | 94.8613 |
| Deep Fade M Egress Floor Close | 2461121.9590 | 15.0 | 0.9187 |

### 3.1 Visual Profile Variations at Deep Minimum

Throughout this long-term monitoring programme, a distinct qualitative shift in the stellar profile of WR 59 has been consistently observed during its deepest fading phases. This structural variation is highly reproducible and has been recorded during every epoch reaching the absolute historical floor of magnitude 15.2. At high magnification with the 22-inch telescope, the normal quiescent state of WR 59 (magnitude 13.9 to 14.0) shows a sharp, pinpoint star identical to the surrounding field stars. However, as the fade progresses and the system touches the magnitude 15.2 floor, the pinpoint core noticeably softens and effectively disappears. Viewed against the nearby magnitude 15.0 comparison star, the target takes on a slightly diffuse, extended appearance. As soon as the system initiates even a minor brightness recovery towards magnitude 15.15, the sharp, distinctive pinpoint stellar core immediately returns.

This rapid transition rules out atmospheric seeing variations, as the comparison star remains consistently sharp throughout the fluctuations. Instead, this visual profile variation reveals a structural change occurring at the minimum floor. At magnitude 15.2, the primary core's direct point-source starlight is heavily attenuated, leaving only a faint halo of scattered light. The instant return of the pinpoint at magnitude 15.15 marks the exact moment direct starlight re-emerges.

### 3.2 The Early Baseline and Initial Deep Drops (Fades A, B, and C)

Before the star began its rapid dust-fading fluctuations, it followed a much slower, highly independent pattern. After establishing a stable four-year baseline at magnitude 13.9, the system broke its long-term stability by executing three distinct, deep plunges down to its absolute minimum thresholds. These early events, designated here as Fades A, B, and C, were major independent episodes because the star fully recovered and returned to its normal quiescent brightness between each fade. Table 1 details the observational records and long intervals for this initial period of the first three deep fades.

#### 3.2.1 Analysis of Fades (A, B, and C)

The first three major events recorded in the campaign behaved similarly, characterised by very deep, clean plunges down to the star's absolute faintest thresholds. Deep Fade A, the historic discovery event on 11 May 2021, saw the star plunge 1.3 magnitudes from its baseline of magnitude 13.9 on JD 2459340 to a minimum of magnitude 15.2 centred at JD 2459346. It remained at this floor for roughly one day before entering a highly structured, slow 14-day recovery egress phase back to the normal quiescent baseline of magnitude 13.9 on JD 2459362 (Fig. 3).

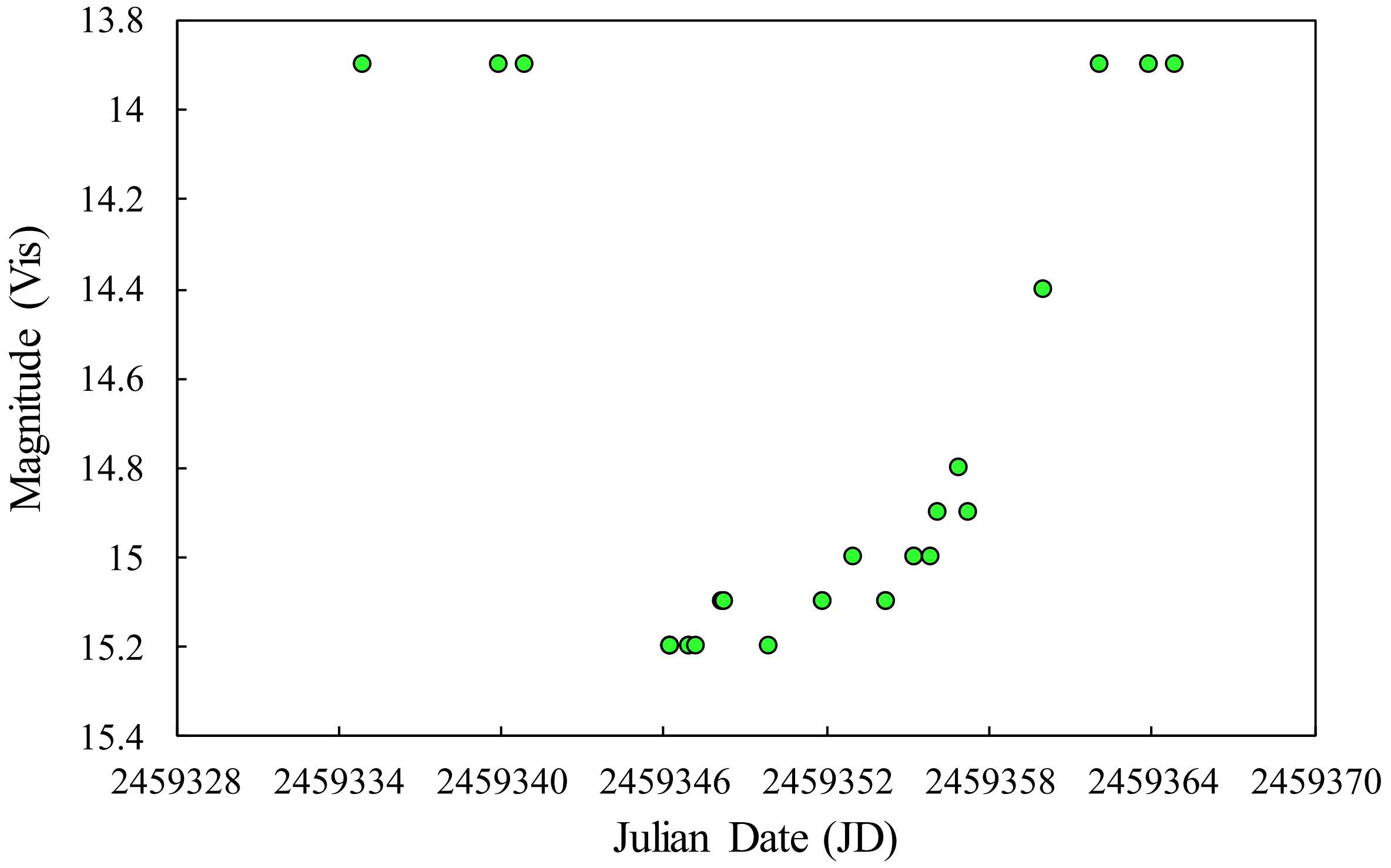

**Figure 3.** High-density unblended visual light curve of Deep Fade A centred on the May 2021 discovery epoch. The plot illustrates the historic profile of the 1.3-magnitude plunge from the stable magnitude 13.9 baseline down to the absolute minimum floor of magnitude 15.2 at JD 2459346, followed by a highly structured, gradual 14-day egress recovery segment returning to the quiescent baseline on JD 2459362. Green circles indicate individual visual magnitude estimates.

Deep Fade B occurred in October 2022 with the minimum centred near JD 2459871, and was characterised by an incredibly rapid ingress phase. Nightly visual estimates caught the star mid-plunge as it dropped cleanly from its magnitude 14.0 baseline at JD 2459860 down through magnitudes 14.4, 14.6, and 15.0 before bottoming out at a minimum of magnitude 15.1. The return to its magnitude 14.0 baseline was observed 15 days later (Fig. 4).

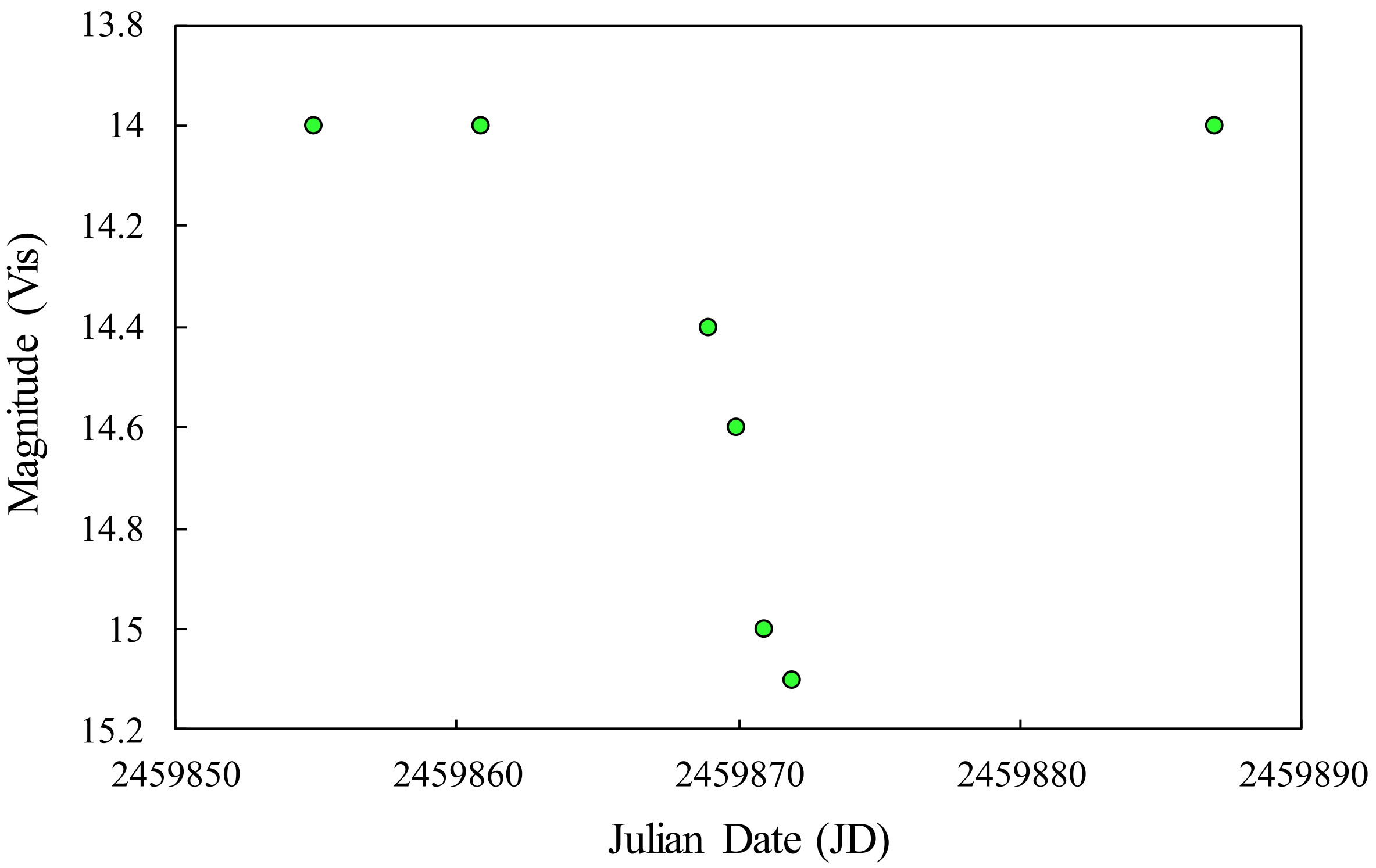

**Figure 4.** High-density unblended light curve of Deep Fade B observed in October 2022 over the Julian Date interval spanning JD 2459854 to JD 2459886. The plot shows an exceptionally steep, rapid ingress phase breaking sharply away from the steady magnitude 14.0 pre-fade baseline. Consecutive nightly visual estimates caught the mid-plunge as it dropped cleanly down through magnitudes 14.4, 14.6, and 15.0 on consecutive nights. The system bottomed out near JD 2459871, reaching a minimum floor of magnitude 15.1 before recovering to its unblended quiescent baseline. Green circles indicate individual visual magnitude estimates.

Deep Fade C followed in August 2023, with the minimum centred near JD 2460169, and displayed a much more turbulent structure than the previous two fades. The leading edge of the ingress showed the star fluctuating between magnitudes 14.4, 14.3, and 14.6 before bottoming out at magnitude 15.1. The subsequent recovery phase was marked by an extended brightening and fluctuating plateau between magnitudes 14.4 and 14.9, where the star lingered for several days before finally breaking free back to its unblended magnitude 14.0 baseline, followed by a 0.1-magnitude variation (Fig. 5).

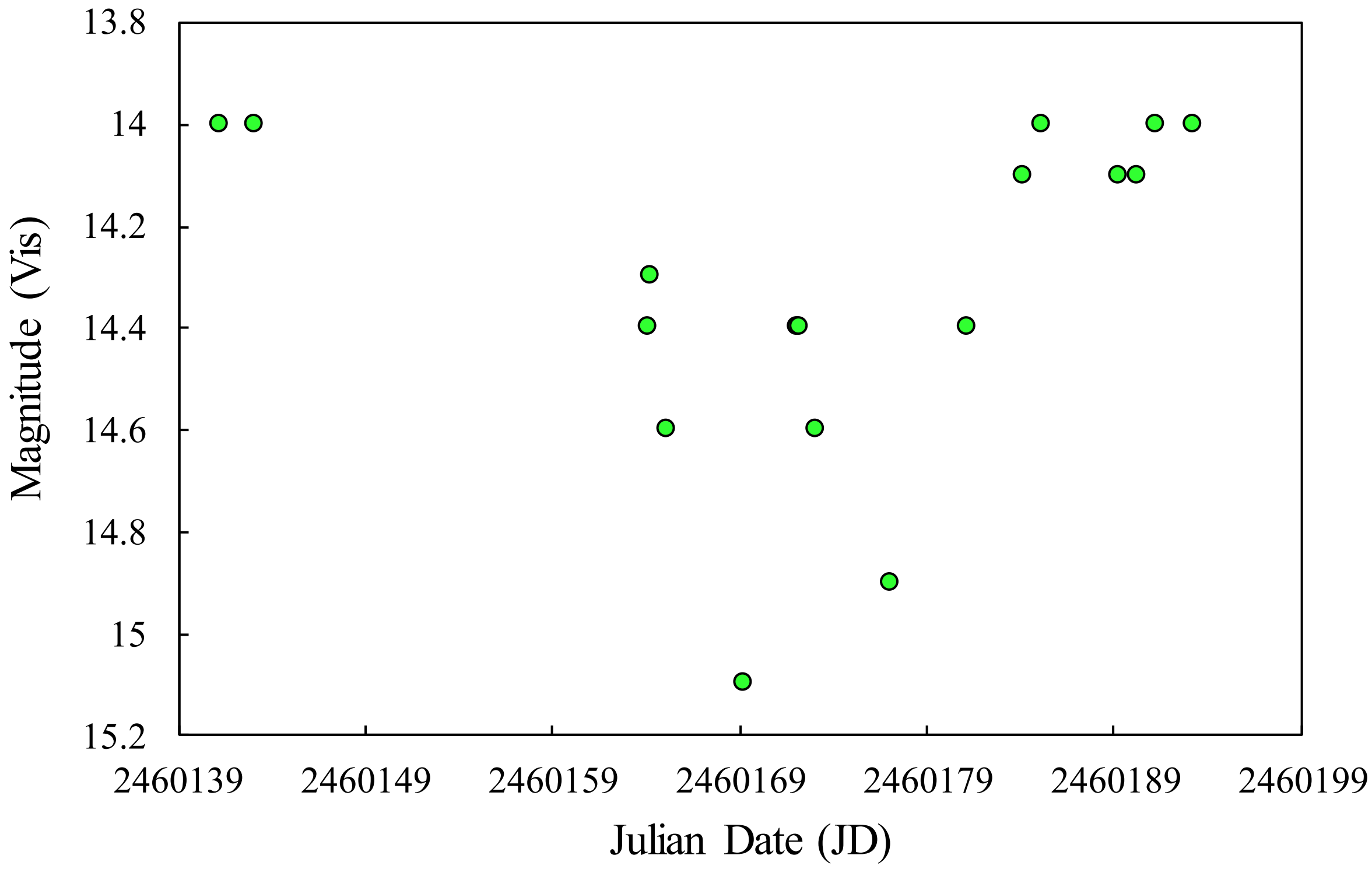

**Figure 5.** High-density unblended light curve for Deep Fade C observed in August 2023 over the Julian Date interval spanning JD 2460141 to JD 2460192. In contrast to the cleaner plunges of the prior two episodes, this profile exhibits a highly turbulent ingress structure, characterised by heavy fluctuations between magnitudes 14.3 and 14.6 along the leading edge of the plunge. The star reaches its absolute minimum floor at magnitude 15.1 on JD 2460169. The subsequent egress recovery is uniquely defined by a broad, flat brightening plateau between magnitudes 14.4 and 14.9, where the system lingered for several days before executing its final clearing back to the unblended magnitude 14.0 baseline. Green circles indicate individual visual magnitude estimates.

### 3.3 The Intermediate Shallow Fade D

Not all observed dust events forced the star down to its absolute minimum thresholds of magnitude 15.1 or 15.2, as documented in Table 2 and shown in Figure 6 during the January 2024 fading event. Instead, the long-term observational record reveals a distinct sequence of intermediate shallow variations in which the star remained noticeably brighter. This behaviour represents a significantly smaller reduction in the star's apparent brightness. Following the earlier deep drops, the star exhibited a distinct shallow profile.

**Table 2. O**bservational log for Intermediate Shallow Fade D

| Observed Event | Julian Date (JD) | Visual Mag | Interval (Days) |
|---|---|---|---|
| Pre-Fade Baseline | 2460322.1510 | 14.0 | Baseline |
| Baseline Confirmation | 2460327.1510 | 14.0 | 5.0000 |
| Ingress Drop | 2460329.1250 | 14.6 | 1.9740 |
| Fade D Minimum | 2460330.1920 | 14.8 | 1.0670 |
| Minimum D Hold | 2460338.1260 | 14.8 | 7.9340 |
| Egress Brightening Step | 2460340.1580 | 14.4 | 2.0320 |
| Baseline Recovery | 2460341.0650 | 14.0 | 0.9070 |

### 3.3.1 Analysis of Fade D

The shallow fading episode of Fade D formed a very distinct profile that departed from the stable baseline of magnitude 14.0 in January 2024 on JD 2460329, where the star dropped quickly to magnitude 14.6. Within approximately one day, it reached its minimum floor of magnitude 14.8 on JD 2460330. A verification observation eight days later on JD 2460338 confirmed the star was still holding at magnitude 14.8. The final recovery was exceptionally rapid, with the star returning completely to its unblended baseline of magnitude 14.0 less than 24 hours later on JD 2460341.

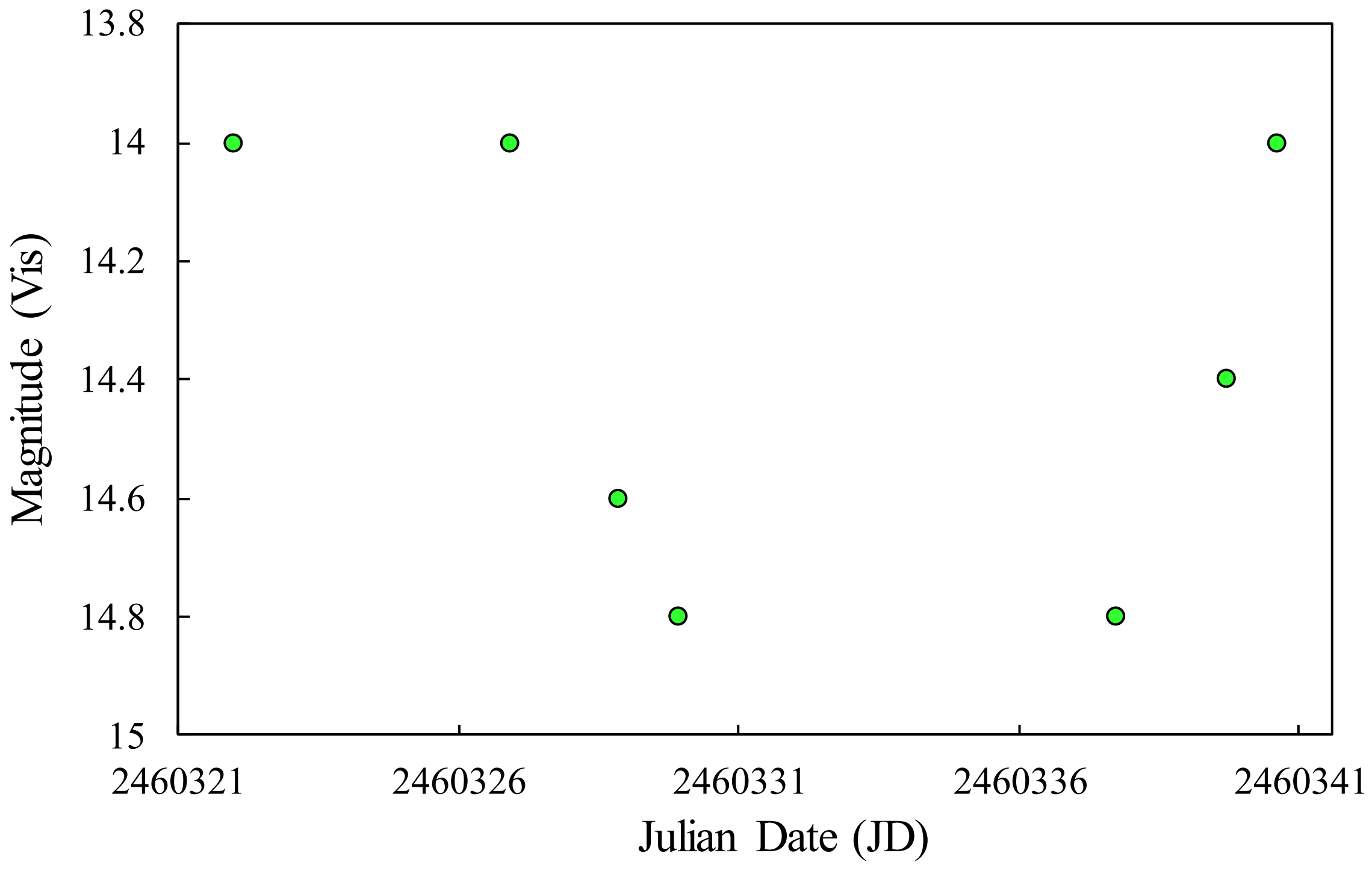

**Figure 6.** High-density unblended light curve of the intermediate shallow Fade D observed in January 2024 over the Julian Date interval spanning JD 2460322 to JD 2460341. The plot displays a sharp, symmetric profile characterised by a rapid drop away from the magnitude 14.0 baseline, an eight-day hold at magnitude 14.8, and a rapid, two-step egress back to the magnitude 14.0 baseline. Green circles indicate individual visual magnitude estimates.

### 3.4 Intermediate Multi-Wave Activity of (Fades E, F, and G)

Following the recovery of the early deep independent events, the system moved into a highly complex phase of intermediate activity from May 2024 to December 2024. As documented in Table 3 and shown in Figure 7, this extended epoch was characterised by multiple dipping waves, rapid multi-night brightness fluctuations, and a prolonged, flat-bottomed deep minimum floor holding steady near magnitude 14.9 during the final wave.

**Table 3.** High-density observational log for the multi-wave campaign (Fades E, F, and G)

| Observed Event | Julian Date (JD) | Visual Mag | Interval (Days) |
|---|---|---|---|
| Post-Fade E Baseline | 2460451.9090 | 14.0 | Baseline |
| Baseline Stability | 2460455.8540 | 14.0 | 3.9450 |
| Pre-Plunge Baseline State | 2460456.8510 | 14.0 | 0.9970 |
| Initial Wave Ingress | 2460457.8760 | 14.6 | 1.0250 |
| Continuous Ingress | 2460457.9150 | 14.6 | 0.0390 |
| Deepening Phase | 2460458.8840 | 14.7 | 0.9690 |
| Near-Minimum | 2460459.9060 | 14.7 | 1.0220 |
| Brightening Shift | 2460462.2700 | 14.3 | 2.3640 |
| First Minimum E | 2460462.8770 | 14.8 | 0.6070 |
| Egress Brightening | 2460494.1450 | 14.0 | 31.2680 |
| Baseline Verification | 2460518.0900 | 14.0 | 23.9450 |
| Second Wave Onset | 2460522.0660 | 14.4 | 3.9760 |
| Small Fluctuation | 2460522.9670 | 14.3 | 0.9010 |
| Secondary Plunge | 2460524.0100 | 14.4 | 1.0430 |
| Minimum Start F | 2460526.0510 | 14.6 | 2.0410 |
| Minimum Hold F | 2460526.1150 | 14.6 | 0.0640 |
| Egress Brightening Shift | 2460527.9900 | 14.4 | 1.8750 |
| Near-Baseline Step | 2460529.9630 | 14.1 | 1.9730 |
| Baseline Stability | 2460536.9400 | 14.0 | 6.9770 |
| Pre-Plunge Anchor | 2460570.8920 | 14.0 | 33.9520 |
| Third Ingress Shift | 2460580.9160 | 14.3 | 10.0240 |
| Secondary Ingress | 2460583.9180 | 14.6 | 3.0020 |
| Deep Minimum G | 2460585.9110 | 14.9 | 1.9930 |
| Intermediate Clearing | 2460586.9870 | 14.8 | 1.0760 |
| Minimum Floor G | 2460590.9080 | 14.9 | 3.9210 |
| Minimum Hold G | 2460592.9140 | 14.9 | 2.0060 |
| Egress Brightening | 2460593.9170 | 14.8 | 1.0030 |
| Strong Recovery | 2460595.9080 | 14.4 | 1.9910 |
| Stabilisation Shelf | 2460603.9120 | 14.4 | 8.0040 |
| Near-Baseline | 2460608.9130 | 14.1 | 5.0010 |
| Continuous Upward | 2460611.9290 | 14.1 | 3.0160 |
| Full Baseline Recovery | 2460646.2260 | 14.0 | 34.2970 |

#### 3.4.1 Analysis and Chronology of Fade E

Fade E represents the initial breakdown of long-term baseline stability following a short, quiescent interval in May 2024 near JD 2460455. The star executed an extremely rapid ingress phase, dropping 0.6 magnitudes from its baseline down to magnitude 14.6 in approximately 24 hours between JD 2460456 and JD 2460457. This steep decline continued to a deepening near-minimum phase of magnitude 14.7 before encountering a brief, pronounced clearing event that temporarily lifted the star back to magnitude 14.3 on JD 2460462.2700.

As soon as this abrupt reversal concluded, the star plunged to its local minimum floor for this episode, hitting magnitude 14.8 on JD 2460462.8770. In contrast to the quick ingress, recovery was highly prolonged and gradual. The star was observed 31 days later at its normal unblended baseline of magnitude 14.0 on JD 2460494.

### 3.4.2 Analysis and Chronology of Fade F

After a quiescent interval lasting nearly 24 days, the system initiated Fade F in July 2024 near JD 2460522. This second intermediate wave was structurally distinct, showing a highly fragmented ingress profile. The onset began with a minor drop to magnitude 14.4, followed by a slight, rapid fluctuation back to magnitude 14.3 before executing a secondary plunge down to magnitude 14.4 on JD 2460524.

The true minimum of this short-duration event reached magnitude 14.6 on JD 2460526, with the star holding flat at this floor for less than two hours before recovering on a steep, rapid egress clearance. The system swept through an egress brightening shift at magnitude 14.4 and a near-baseline step at magnitude 14.1, sealing a complete return to baseline stability at magnitude 14.0 by JD 2460536. The entire profile of Fade F was compact, lasting approximately 15 days from initial onset to baseline recovery.

### 3.4.3 Analysis and Chronology of Fade G

The final and deepest component of this multi-wave cluster emerged after an extended, nearly 34-day pre-plunge anchor phase. The onset of this dimming was detected in September 2024 near JD 2460580, with the star sliding through magnitude 14.3 down into a secondary ingress wall at magnitude 14.6. The system abruptly entered its deep minimum floor at magnitude 14.9 on JD 2460585.

Fade G exhibited a highly complex, double-bottomed minimum structure. After touching magnitude 14.9, the star experienced a brief 1-day intermediate clearing up to magnitude 14.8, before sliding right back down into a prolonged, two-day minimum hold at magnitude 14.9 from JD 2460590 to JD 2460592. The initial egress brightening began on JD 2460593, climbing through a sequence of strong recovery phases and temporary stabilisation shelves at magnitude 14.4 before permanently securing its full unblended baseline of magnitude 14.0 on JD 2460646.

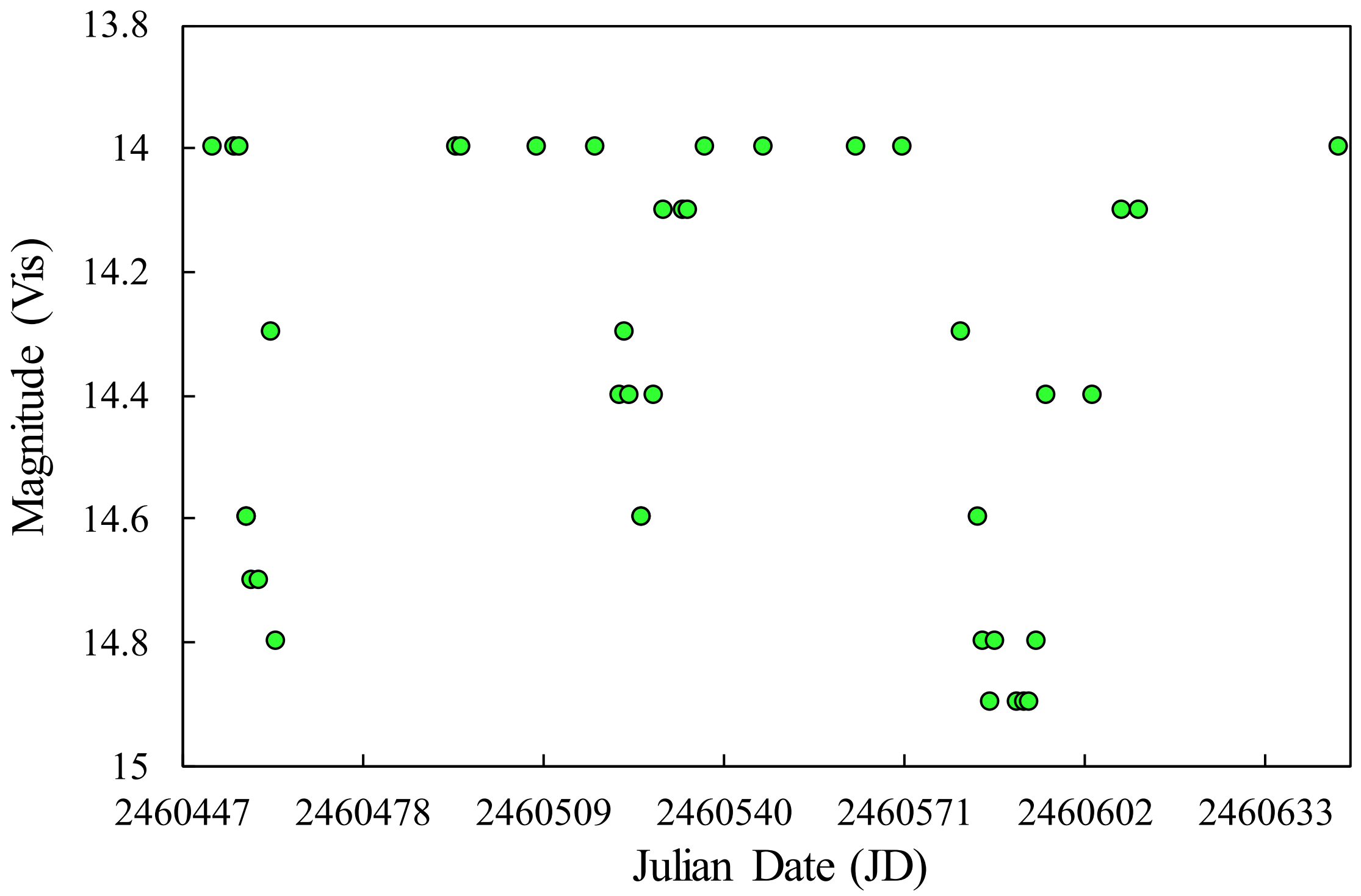

**Figure 7.** High-density unblended light curve illustrating the complex intermediate multi-wave epoch of Fades E, F, and G, over the Julian Date interval spanning JD 2460451 to JD 2460646. This highly dynamic phase is characterised by three sequential dipping waves separated by short, stable intervals. The initial component Fade E exhibits an exceptionally rapid ingress near JD 2460457 down to magnitude 14.6, encountering a clearing at magnitude 14.3 before bottoming out at its true event minimum of magnitude 14.8 on JD 2460462.8770. The secondary wave Fade F forms a compact, 15-day sequence centred at a sharp core dip of magnitude 14.6 on JD 2460526. The final and deepest component Fade G features a rapid descent to magnitude 14.9 on JD 2460585, displaying a complex minimum floor variation with a prolonged two-day hold before climbing through temporary stabilisation shelves at magnitude 14.4, 14.1, and 14.0 to secure a full baseline recovery. Green circles indicate individual visual magnitude estimates.

### 3.5 The Pre-Active Epoch Fluctuations of (Fades H, I, J, K, and L)

Following the more prolonged events, the star entered a long, unsettled state characterised by multiple brief, short-lived dips from March 2025 to January 2026. As documented in Table 4 and shown in Figure 8, this sequence represents a volatile phase where the star repeatedly fluctuated downward to intermediate limits near magnitudes 14.4 and 14.6 before quickly executing swift baseline returns. These rapid, undulating waves were separated by extended, weeks-long periods of flat baseline stability.

**Table 4.** High-density **o**bservational log for pre-active fluctuations of (Fades H, I, J, K, and L)

| Observed Event | Julian Date (JD) | Visual Mag | Interval (Days) |
|---|---|---|---|
| Baseline Opening | 2460754.0800 | 14.0 | Baseline |
| Baseline Stability | 2460757.1170 | 14.0 | 3.0370 |
| Flat Wave Minimum H | 2460760.0410 | 14.4 | 2.9240 |
| Hold at Dimmed Limit H | 2460775.8970 | 14.4 | 15.8560 |
| Baseline Recovery | 2460781.9130 | 14.0 | 6.0160 |
| Pre-Drop Baseline | 2460815.9150 | 14.0 | 34.0020 |
| Secondary Sharp Dip I | 2460820.9010 | 14.4 | 4.9860 |
| Swift Baseline Return | 2460827.1460 | 14.0 | 6.2450 |
| Long Baseline Plateau | 2460874.0790 | 14.0 | 46.9330 |
| Third Downward Plunge J | 2460879.0250 | 14.4 | 4.9460 |
| Near-Floor Brightening | 2460880.9700 | 14.3 | 1.9450 |
| Flat Wave Minimum | 2460887.0340 | 14.4 | 6.0640 |
| Flat Wave Minimum Hold J | 2460891.0430 | 14.4 | 4.0090 |
| Rapid Recovery Rise | 2460895.9490 | 14.0 | 4.9060 |
| Long-Term Baseline Hold | 2460911.9690 | 14.0 | 16.0200 |
| Fourth Ingress Shift | 2460932.9542 | 14.4 | 20.9852 |
| Sustained Minimum State | 2460933.8930 | 14.4 | 0.9388 |
| Deepest Minimum K | 2460935.9060 | 14.6 | 2.0130 |
| Wave Egress Step | 2460936.9890 | 14.4 | 1.0830 |
| Upward Gradient Step | 2460946.9290 | 14.1 | 9.9400 |
| Full Baseline Recovery | 2460961.9330 | 14.0 | 15.0040 |
| Pre-Drop Stable Baseline | 2461020.1720 | 14.0 | 58.2390 |
| Main Fade Minimum L | 2461026.1790 | 14.6 | 6.0070 |
| Baseline Stability | 2461063.0979 | 14.1 | 36.9189 |

### 3.5.1 Analysis of (Fades H, I, J, K, and L)

Analysis of the observational log reveals that the system experienced multiple brief dipping structures separated by weeks-long periods of flat baseline stability. The initial drop Fade H triggered near JD 2460757, with the star slipping down to a flat wave minimum of magnitude 14.4 on JD 2460760. The system held this dimmed state for a prolonged 15.8560 days before initiating a 6-day clearing back to the magnitude 14.0 baseline. Following a 34-day stable interval, a second nearly identical drop Fade I occurred near JD 2460820, plunging swiftly to magnitude 14.4 before executing an equally rapid 6.2-day return to the baseline of magnitude 14.0.

A third major undulating sequence Fade J initiated near JD 2460879 after a long 46.9-day plateau. This event displayed a multi-stage profile, spanning over a week in total duration. A brief egress brightening to magnitude 14.3 occurred on JD 2460880 followed by an intermediate floor at magnitude 14.4 to JD 2460891, before recovering cleanly to the magnitude 14.0 baseline on JD 2460895. This systematic pattern repeated a fourth time near JD 2460932 with Fade K, forcing a deeper local threshold drop through magnitude 14.4 down to a local minimum of magnitude 14.6 on JD 2460935, followed by a structured, multi-step egress gradient rise to a baseline of magnitude 14.0 on JD 2460961.

Following an extended 58-day baseline stability, the final independent drop of this sequence Fade L triggered near JD 2461020. The star plummeted cleanly to its main minimum at magnitude 14.6 on JD 2461026. Crucially, the egress from Fade L did not achieve a full recovery back to the magnitude 14.0 baseline. Instead, the recovery stalled completely, remaining fixed on a prolonged sub-baseline plateau at magnitude 14.1 for over 36 days. This unrecovered shelf served as the final shallow fade observational indicator before the system transitioned directly into its active dust-production epoch.

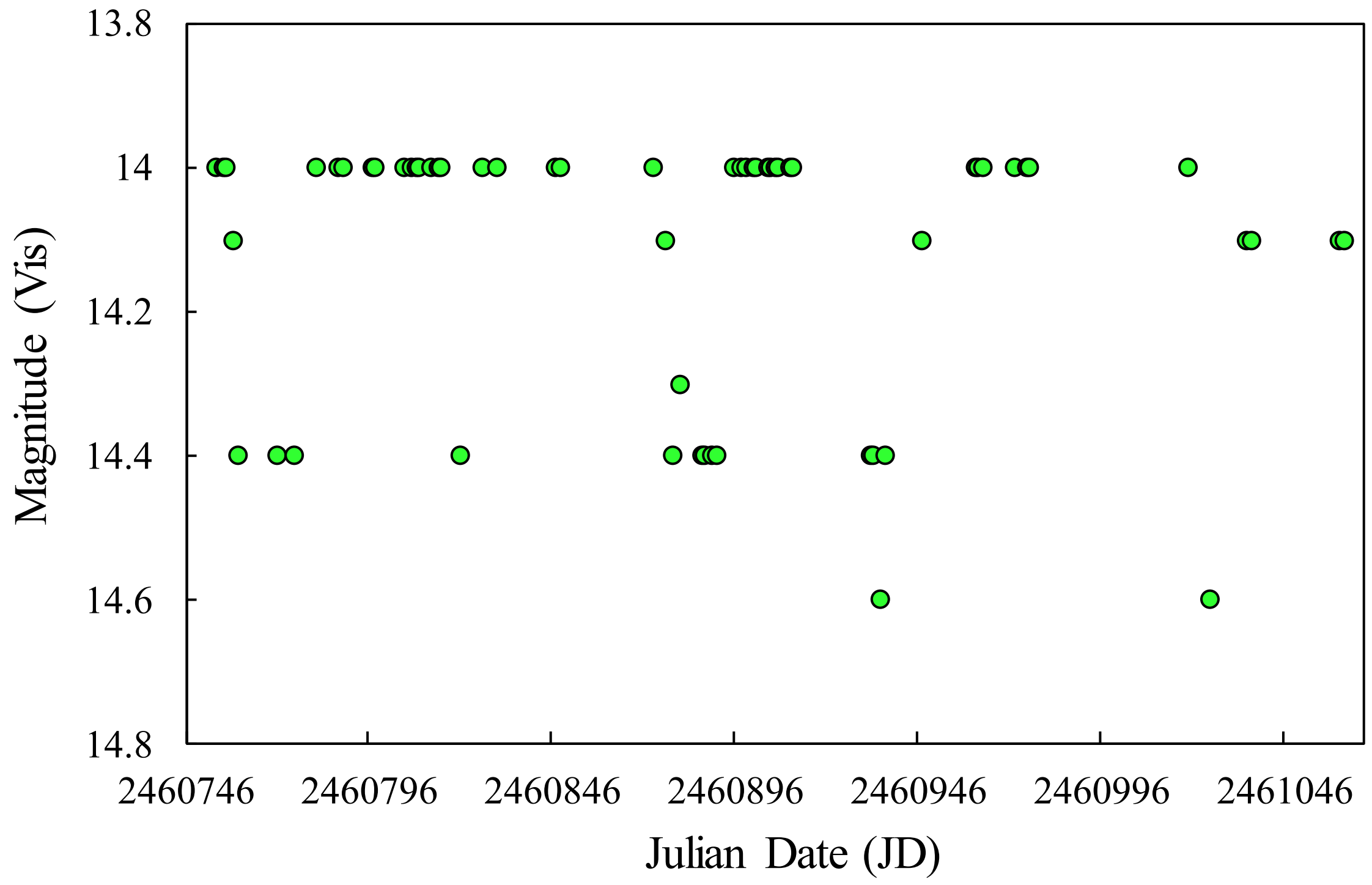

**Figure 8.** Consolidated high-density light curve of the pre-active epoch fluctuations of Fades H, I, J, K, and L, over the Julian Date interval spanning JD 2460754 to JD 2461063. This volatile sequence consists of five rapid, intermediate dipping sequences separated by prolonged periods of flat baseline stability. The initial event Fade H drops cleanly to a flat wave minimum of magnitude 14.4 on JD 2460760, holding for 15.8560 days. Following a 34-day interval, a nearly identical sharp plunge Fade I triggers near JD 2460820, reaching magnitude 14.4 before a rapid 6.2-day recovery. The third sequence Fade J initiates near JD 2460879, exhibiting a multi-stage visual profile and spanning over a week in total duration. A brief egress brightening to magnitude 14.3 occurred on JD 2460880, followed by an intermediate floor at magnitude 14.4 to JD 2460891, before recovering cleanly to the magnitude 14.0 baseline on JD 2460895. This pattern repeats a fourth time near JD 2460932 Fade K, forcing a deeper local minimum of magnitude 14.6 on JD 2460935, followed by a multi-step egress. The final independent event Fade L plunges to a minimum of magnitude 14.6 on JD 2461026. Crucially, the egress fails to achieve a full baseline recovery, stalling on a prolonged sub-baseline plateau at magnitude 14.1 for over 36 days. Green circles indicate individual visual magnitude estimates.

### 3.6 Symmetrical Intermediate Activity of Fade M

After a short period of baseline quiescence, the star executed a highly distinct, symmetrical intermediate plunge in March 2026 designated here as Fade M. As documented in Table 5 and Figure 9, this event broke away from a temporary magnitude 14.1 sub-baseline plateau and underwent a rapid, clean ingress drop down to a flat floor minimum at magnitude 15.0. The subsequent recovery mirrored the descent velocity almost perfectly, climbing steadily on consecutive nights back to the baseline of magnitude 14.0.

**Table 5.** High-density observational log for Fade M

| Observed Event | Julian Date (JD) | Visual Mag | Interval (Days) |
|---|---|---|---|
| Pre-Baseline Brightening | 2461090.0597 | 14.0 | Baseline |
| Sub-Baseline Plateau | 2461095.1222 | 14.1 | 5.0625 |
| Monitoring Baseline Anchor | 2461105.1361 | 14.1 | 10.0139 |
| Pre-Plunge Hold | 2461109.0299 | 14.1 | 3.8938 |
| Main Ingress Drop | 2461113.0979 | 14.8 | 4.0680 |
| Floor Minimum Reach M | 2461121.0403 | 15.0 | 7.9424 |
| Floor Minimum Hold M | 2461121.9590 | 15.0 | 0.9187 |
| Initial Egress Brightening | 2461122.0646 | 14.9 | 0.1056 |
| Continuous Recovery Trend | 2461122.9465 | 14.6 | 0.8819 |
| Recovery Stabilisation Plateau | 2461123.0389 | 14.6 | 0.0924 |
| Accelerated Brightening | 2461124.0639 | 14.4 | 1.0250 |
| Near-Baseline Anchor | 2461129.9903 | 14.1 | 5.9264 |
| Full Baseline Recovery | 2461134.9215 | 14.0 | 4.9312 |

#### 3.6.1 Analysis and Chronology of Fade M

Analysis of the observational log reveals that the star had a pre-baseline brightening and then held steady near magnitude 14.1 through JD 2461109, before experiencing a sudden, sharp plunge down to magnitude 14.8 on JD 2461113. It reached its absolute minimum floor of magnitude 15.0 on JD 2461121, and remained pinned flat at this minimum for approximately one day.

The recovery phase was highly symmetrical to the ingress. The star broke free from the floor on JD 2461122.0646 and climbed steadily on consecutive nights through magnitudes 14.9, 14.6, and 14.4 on JD 2461124. After a brief pause near magnitude 14.1, it completed a full recovery back to the baseline of magnitude 14.0 on JD 2461134. This clean sequence represents the last symmetrical, deep independent fade before the star entered its 2026 continuous multi-plunge, unstable state.

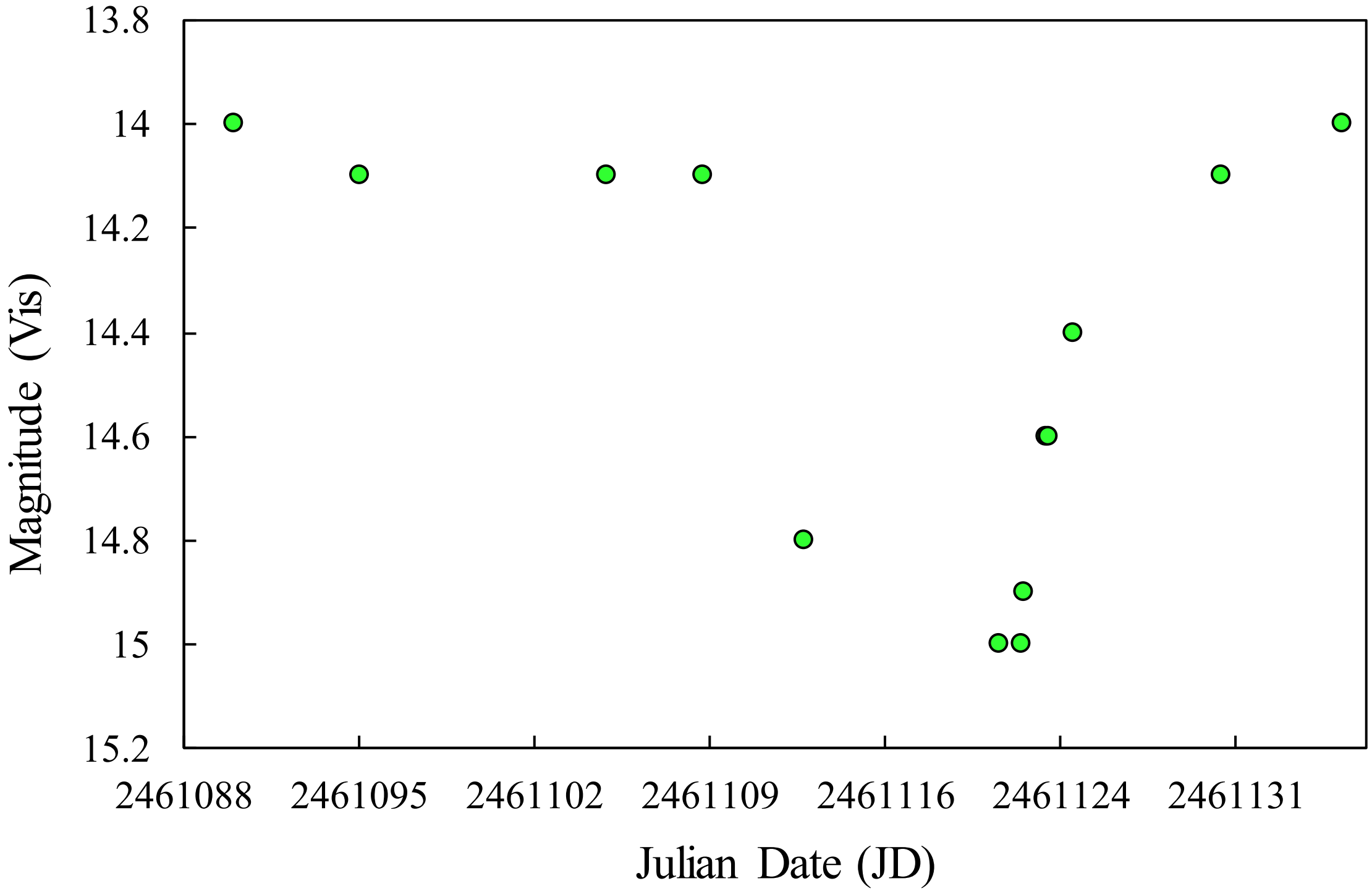

**Figure 9.** High-density unblended light curve of the symmetrical intermediate fading event, designated here as Fade M, over the Julian Date interval spanning JD 2461090 to JD 2461134. The profile highlights a highly distinct, visual plunge breaking away from a magnitude 14.1 sub-baseline plateau. The system undergoes a rapid, clean ingress drop down to magnitude 14.8 on JD 2461113, before bottoming out at an absolute floor minimum of magnitude 15.0 on JD 2461121, remaining pinned flat for approximately 22 hours. The subsequent egress phase mirrors the descent velocity almost perfectly, climbing steadily on consecutive nights through magnitudes 14.9, 14.6, and 14.4. Following a temporary stabilisation plateau near magnitude 14.1, the star completes a full recovery back to its baseline of magnitude 14.0 on JD 2461134, marking the final independent, symmetrical sequence prior to the system's unstable epoch. Green circles indicate individual visual magnitude estimates.

### 3.7 The Master Visual Record and the 2026 Continuous Fade Era

Following the brief recovery of Fade M, the system entered its most chaotic and prolonged state of visual obscuration. To establish the long-term physical context of this transition, a comprehensive master visual light curve is presented in Figure 10, compiled entirely from pure, unblended magnitude estimates across the full campaign window spanning April 2017 (JD 2457867) to November 2026 (JD 2461362). This visual dataset cleanly highlights the system's long-term behavioural evolution, completely free from contamination introduced by automated survey pipelines.

The compiled observational record frames a distinct structural shift in the lifecycle of WR 59. The long-term record opens with an initial four-year quiescent plateau at a baseline near magnitude 13.9, representing a prolonged era of quiescent stability. This baseline is broken by the historic May 2021 plunge of Deep Fade A to magnitude 15.2, followed by the subsequent independent deep minima of Fades B and C. As the monitoring campaign transitioned into the mid-2020s, the visual log documents a progressive compression of inter-fade intervals, recording the system through intermediate multi-wave clusters and fluctuations. The master dataset culminates in an unprecedented, high-density near-vertical cluster of data entries at the extreme right of the plot. This feature captures the star's total transition as it completely abandons its historical seasonal baselines to enter the current ongoing 2026 continuous Fade N complex epoch.

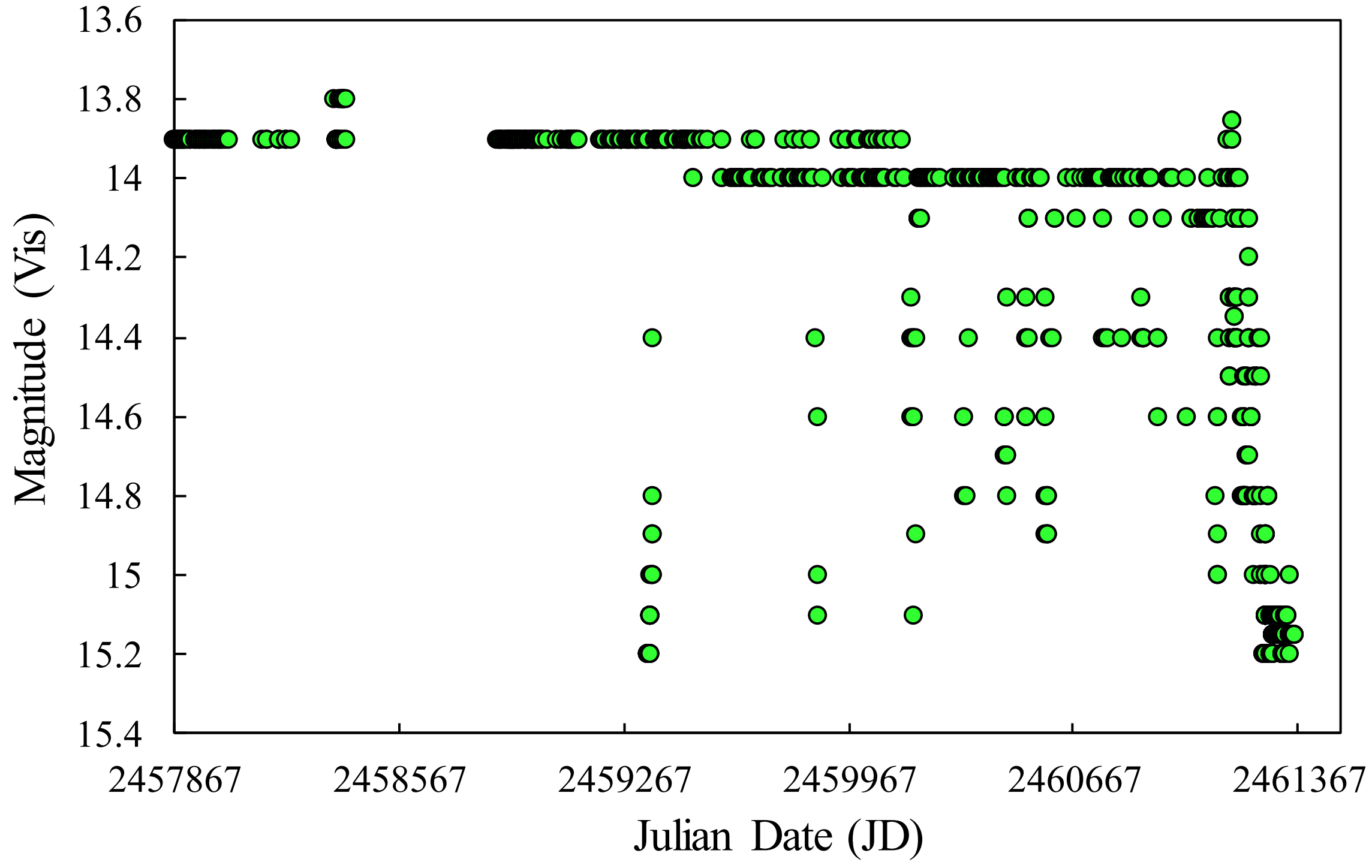

**Figure 10.** Comprehensive master light curve of WR 59 compiled from pure, unblended magnitude estimates across the full campaign interval spanning from April 2017 (JD 2457867) to November 2026 (JD 2461362). The observational record frames the initial four-year quiescent plateau at a baseline near magnitude 13.9, the observational gap and the historic May 2021 plunge of Deep Fade A to magnitude 15.2. As the campaign transitions into the mid-2020s, the plot documents the progressive compression of inter-fade intervals, mapping the system through the multi-wave clusters. The unprecedented high-density cluster of data points at the extreme right of the plot records the star's sustained profile transition as it abandons its seasonal baselines to enter the ongoing 2026 continuous Fade N complex epoch. Green circles indicate individual visual magnitude estimates.

### 3.8 High-Density Structure of the Fade N Complex

My ground-based visual monitoring of WR 59 across a 228-day interval from April 2026 (JD 2461134) to November 2026 (JD 2461362) documented a profound, multi-tiered transition away from the star's historical baseline state. The light curve as shown in Figure 11 documents a complex, staircase descent profile rather than a smooth, uniform geometric decline. During the initial descent phase (JD 2461134 to 2461210), I recorded severe instability, noting large-amplitude variations where the star bounced rapidly between magnitude 13.9 and 14.5. Following a brief stabilisation near magnitude 14.8, I recorded a rapid, pronounced brightening event near JD 2461219, where the system recovered back to magnitude 14.1 before undergoing its final, step-descent.

After JD 2461259, the star entered a prolonged, dense minimum floor, flattening out between magnitude 15.1 and 15.2. To monitor this deep phase, I anchored my estimates against a single, stable magnitude 15.0 pinpoint comparison star in the immediate field. By using this close-field comparison technique, I could reliably estimate the deeper drops as the star faded in and out of view, virtually disappearing at magnitude 15.2 relative to my fixed magnitude 15.0 star. The persistent clustering of my data points documents that even when the star is most heavily obscured, the system remains intensely active, experiencing rapid, localised micro-variations.

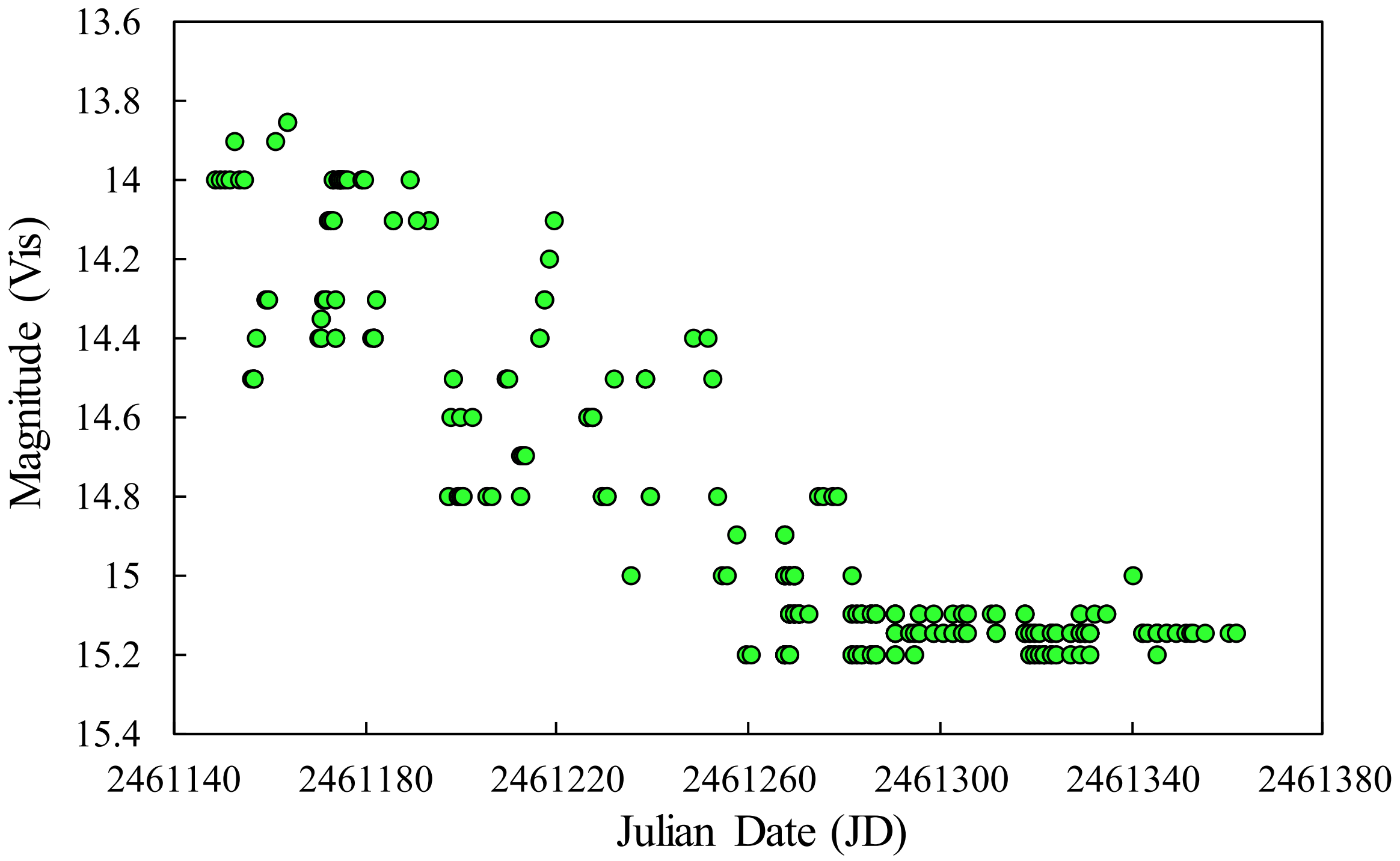

**Figure 11.** High-density light curve illustrating the structure of the ongoing Fade N complex over the Julian Date interval spanning JD 2461134 to JD 2461362. This zoomed-in perspective resolves the highly volatile staircase profile of the 2026 major obscuration epoch, displaying a high density of nightly observations made entirely at the eyepiece. The observational record begins with rapid, short-lived initial drops that bounce between magnitude 14.0 and magnitude 14.5, before the system completely abandons its independent oscillating behaviour near JD 2461181. The star enters a locked, non-recovering downward step-descent, characterised by a series of failed recovery steps that remain fixed along extended sub-baseline plateaus at magnitude 14.1 and 14.8. The final step-descent forces the pinpoint stellar core straight down into its absolute historical minimum floor of magnitude 15.2. At the far right of the plot, the star remains heavily obscured, showing variations at this deep limit for a sustained 102 consecutive days up to the close of the logged campaign. Green circles indicate individual visual magnitude estimates.

### 3.8.1 Chronological Logs and Analysis of the Initial Descent Fade N Complex (Part 1)

The high-density chronological monitoring compiled across this Julian Date complex interval (Tables 6a and 6b) documents the system through its rapid initial descent phases before the deep core attenuation. Because this major obscuration event is currently ongoing with no end in sight, these chronological log tables conclude at JD 2461286, while the broader Fade N complex light curve (Fig. 11) continues to record the minimum floor. Segments of this ongoing deep fade are covered in detail by the high-density intra-night campaigns that follow.

**Table 6a.** Observational log for the modern active phase (Part 1: The Initial Descent)

| Observed Event | Julian Date (JD) | Visual Mag | Interval (Days) |
|---|---|---|---|
| Full Baseline Return | 2461134.9215 | 14.0 | *Session Open* |
| Baseline Stability State | 2461149.0285 | 14.0 | 14.1070 |
| Peak Brightness Shift | 2461153.0910 | 13.9 | 4.0625 |
| Pre-Drop Baseline State | 2461155.0639 | 14.0 | 1.9729 |
| Minimum Ingress Phase | 2461156.8813 | 14.5 | 1.8174 |
| Minimum Ingress Hold | 2461156.9161 | 14.5 | 0.0348 |
| Egress Brightening | 2461157.8882 | 14.4 | 0.9721 |
| Egress Brightening Step | 2461159.8813 | 14.3 | 1.9931 |
| Baseline Recovery | 2461160.0271 | 14.1 | 0.1458 |
| Full Baseline State | 2461161.8819 | 14.0 | 1.8548 |
| Historical Maximum Peak | 2461164.1618 | 13.85 | 2.2799 |
| Maximum Peak Hold | 2461164.2243 | 13.85 | 0.0625 |
| Sudden Secondary Descent | 2461170.9410 | 14.4 | 6.7167 |
| Secondary Descent Hold | 2461171.1444 | 14.4 | 0.2034 |
| Brightening Shift | 2461171.1951 | 14.35 | 0.0507 |
| Sub-Baseline Fluctuation | 2461171.8743 | 14.3 | 0.6792 |
| Sub-Baseline Hold | 2461171.9993 | 14.3 | 0.1250 |
| Baseline Return | 2461172.8778 | 14.1 | 0.8785 |
| Plateau Hold | 2461173.1326 | 14.1 | 0.2548 |
| Continued Plateau Hold | 2461173.1701 | 14.1 | 0.0375 |
| Full Baseline State | 2461173.8806 | 14.0 | 0.7105 |
| Emerging Ingress Onset | 2461174.0389 | 14.3 | 0.1583 |
| Ingress Minimum | 2461174.1479 | 14.4 | 0.1090 |
| Rapid Egress Recovery | 2461174.8806 | 14.0 | 0.7327 |
| Stable Baseline Return | 2461179.9861 | 14.0 | 5.1055 |

Analysis of the comprehensive observational log across this initial phase reveals a baseline return at magnitude 14.0 on JD 2461134. The system then held a quiet, steady baseline for a two-week period. During this time, a brief brightening event was documented at magnitude 13.9 on JD 2461153, followed later by an absolute maximum peak at magnitude 13.85 on JD 2461164, marking the absolute brightest state of the star during this entire epoch before a series of rapid, multi-day dips commenced.

The first distinct visual departure from stability triggered on JD 2461156, when the star underwent a sudden ingress drop down to magnitude 14.5. High-density observing sessions documented an intra-night hold at this dimmed limit before a rapid egress sweep initiated a full recovery back to the magnitude 14.0 baseline within five days. This quiet state was short-lived. On JD 2461170, a sudden secondary plunge caught the star mid-plunge, driving it back down to magnitude 14.4. The subsequent egress structure was highly variable, showing a brightening shift at magnitude 14.35 and a sub-baseline fluctuation at magnitude 14.3 before the star completely recovered to its baseline.

A final, fleeting drop was observed on JD 2461174.1, dipping the target briefly to magnitude 14.4 before an exceptionally fast recovery returned the star to its baseline on JD 2461174.8. The final baseline confirmation on JD 2461179 represents the absolute close of the star's independent, oscillating behaviour.

The short duration of these dips, often lasting less than 24 hours to a few days from ingress to full egress, demonstrates that the visual data reflect a rapid succession of short-duration, high-frequency brightness variations, causing the star to fluctuate heavily before making its final, temporary baseline recovery.

### 3.8.2 Chronological Logs and Analysis of the Descent Fade N Complex (Part 2)

Following the final brief baseline recovery, the star completely abandoned its independent oscillating pattern and entered a locked, non-recovering downward descent. Table 6b details this severe deep core descent and the subsequent historical floor hold.

**Table 6b.** Observational log for the modern active phase (Part 2: The Main Ingress Descent and Deep Floor Hold)

| Observed Event | Julian Date (JD) | Visual Mag | Interval (Days) |
|---|---|---|---|
| Initial Ingress Phase | 2461181.9750 | 14.4 | Session Open |
| Sustained Ingress Phase | 2461182.0819 | 14.4 | 0.1069 |
| Sustained Ingress Phase | 2461182.2201 | 14.4 | 0.1382 |
| Brightening Step | 2461182.8507 | 14.3 | 0.6306 |
| Extended Egress Brightening | 2461186.1208 | 14.1 | 3.2701 |
| Baseline Brightening Hold | 2461189.9069 | 14.0 | 3.7861 |
| Secondary Ingress Shift | 2461193.9579 | 14.1 | 4.0510 |
| Major Ingress Descent | 2461197.8576 | 14.8 | 3.8997 |
| Brightening Phase | 2461198.0556 | 14.6 | 0.1980 |
| Brightening Peak | 2461198.8681 | 14.5 | 0.8125 |
| Minimum Floor Descent | 2461199.8868 | 14.8 | 1.0187 |
| Brightening Step | 2461200.1940 | 14.6 | 0.3072 |
| Minimum Floor Descent | 2461205.9604 | 14.8 | 5.7664 |
| Brief Brightening Shift | 2461210.0056 | 14.5 | 4.0452 |
| Minimum Floor Descent | 2461212.8597 | 14.8 | 2.8541 |
| Brightening Step | 2461213.1431 | 14.7 | 0.2834 |
| Extended Ingress Brightening | 2461216.8667 | 14.4 | 3.7236 |
| Pre-Floor Brightening Plateau | 2461219.8757 | 14.1 | 3.0090 |
| Ingress Phase | 2461226.8604 | 14.6 | 6.9847 |
| Minimum Floor Descent | 2461230.8875 | 14.8 | 4.0271 |
| Brightening Shift | 2461232.1493 | 14.5 | 1.2618 |

| **Observed Event** | **Julian Date (JD)** | **Visual Mag** | **Interval (Days)** |
|---|---|---|---|
| First Descent to Deep Minimum | 2461236.0986 | 15.0 | 3.9493 |
| Intermediate Brightening Plateau | 2461238.8729 | 14.5 | 2.7743 |
| Sustained Plateau Hold | 2461239.1167 | 14.5 | 0.2438 |
| Minimum Floor Descent | 2461239.9587 | 14.8 | 0.8420 |
| Sustained Minimum Hold | 2461253.8806 | 14.8 | 13.9219 |
| Deep Ingress Phase | 2461254.8792 | 15.0 | 0.9986 |
| Brightening Phase | 2461257.8681 | 14.9 | 2.9889 |
| Historical Minimum Floor | 2461259.8708 | 15.2 | 2.0027 |
| Brightening Phase | 2461267.8361 | 15.0 | 7.9653 |
| Minimum Floor | 2461267.8965 | 15.2 | 0.0604 |
| Brightening Peak | 2461267.9292 | 14.9 | 0.0327 |
| Minimum Floor | 2461267.9375 | 15.2 | 0.0083 |
| Sustained Minimum Floor | 2461268.8965 | 15.2 | 0.9590 |
| Brightening Phase | 2461268.9028 | 15.0 | 0.0063 |
| Minimum Descent | 2461268.9111 | 15.1 | 0.0083 |
| Brightening Phase | 2461269.0007 | 15.0 | 0.0896 |
| Sustained Brightening Phase | 2461269.9222 | 15.0 | 0.9215 |
| Minimum Descent | 2461270.8813 | 15.1 | 0.9591 |
| Sustained Minimum Floor | 2461273.1917 | 15.1 | 2.3104 |
| Brightening Plateau | 2461278.8861 | 14.8 | 5.6944 |
| Total Minimum Floor | 2461281.8764 | 15.2 | 2.9903 |
| Sustained Minimum Floor Close | 2461286.8972 | 15.2 | 5.0208 |

Analysis of this second phase reveals a total shift in the system's behaviour, marked by a relentless staircase-style descent where every attempt at recovery failed. The main ingress descent officially triggered on JD 2461181, with the star dropping from its baseline level down to magnitude 14.4. Unlike the previous short dips, the star could not break free; it managed only a brief breaking egress step to magnitude 14.3 and a short-lived return to magnitude 14.0 on JD 2461189, before moving to a sub-baseline shelf at magnitude 14.1 as the downward plunge resumed in earnest. On JD 2461197, the star underwent a secondary major drop that forced it down to magnitude 14.8. For the next month, the star remained trapped in a highly turbulent state, constantly fluctuating between magnitudes 14.5 and 14.8, documenting a heavily attenuated light profile. A massive push downward on JD 2461236 forced the star to magnitude 15.0 for the first time, and despite a temporary bounce back to magnitude 14.5, the system completely lost outward momentum.

The final plunge occurred on JD 2461259.8708, driving the star straight down to its absolute historical minimum floor of magnitude 15.2. The star has remained heavily obscured ever since, at this varying deep limit for a sustained 27 consecutive days up to the close of this specific log segment at JD 2461286, counting toward an ongoing 102-day minimum hold across the broader campaign. The high-density observing sessions during this deep hold documented brief moments where the target brightened slightly to magnitudes 15.1 and 15.0 and a temporary peak of magnitude 14.8, before immediately sinking back to magnitude 15.2.

Although the direct point-source starlight was heavily obscured, the system remained subject to rapid, short-term micro-variations.

## 4. High-Density Intra-Night Monitoring at Minimum

To investigate short-term variations while the star was faint, I conducted intense, high-density visual monitoring sessions across consecutive nights through JD 2461267 through JD 2461270. These intra-night sessions demonstrate that the star does not sit perfectly still at the bottom of its plunge. Instead, the structure of the heavily attenuated starlight alters dynamically, presenting four distinct visual profiles.

### 4.1 Initial Monitoring Campaign (Nights 1 to 4: JD 2461267–2461270)

To establish a high-resolution baseline of the short-term fluctuations occurring right at the floor of the fade, the initial phase of the intensive monitoring campaign was executed across four consecutive nights. These continuous sessions were designed to record the rapid light variations on minute-by-minute timescales, observing the volatile transitions between flat plateau states and bimodal magnitude oscillations.

#### 4.1.1 Night 1 (JD 2461267): Extreme Bimodal Volatility

**Table 7.** High-density intra-night observational log for Night 1 (JD 2461267)

| **Observed Event** | **Julian Date (JD)** | **Visual Mag** | **Interval (Days)** |
|---|---|---|---|
| Session Initiation | 2461267.8778 | 15.0 | *Session Open* |
| Minimum Floor Verification | 2461267.8965 | 15.2 | 0.0187 |
| Variability Phase | 2461267.9153 | 15.0 | 0.0188 |
| Minimum Hold | 2461267.9229 | 15.2 | 0.0076 |
| Brightening Phase | 2461267.9292 | 14.9 | 0.0063 |
| Minimum Hold | 2461267.9319 | 15.2 | 0.0027 |
| Brightening Phase | 2461267.9347 | 14.9 | 0.0028 |
| Minimum Hold | 2461267.9375 | 15.2 | 0.0028 |
| Variability Phase | 2461267.9514 | 15.0 | 0.0139 |
| Sustained Minimum Core | 2461267.9542 | 15.2 | 0.0028 |
| Minimum Floor Verification | 2461267.9556 | 15.2 | 0.0014 |
| Brightening Phase | 2461267.9583 | 14.9 | 0.0027 |
| Variability Phase | 2461267.9597 | 15.0 | 0.0014 |
| Minimum Floor / Session Close | 2461267.9653 | 15.2 | 0.0056 |

The first intensive intra-night session, logged over a 2.1-hour monitoring window, revealed an unprecedented level of short-term volatility right at the system's absolute historical floor. As documented in Table 7 and Figure 12, the session opened on JD 2461267.8778 with the target holding a slightly softened profile at magnitude 15.0. Within roughly 21 minutes, the system executed a steep plunge down to the absolute minimum floor of magnitude 15.2. Rather than stabilising at this deep limit, the star immediately entered a state of rapid, high-amplitude variations, bouncing erratically between magnitudes 15.2, 15.0, and a peak partial clearance at magnitude 14.9.

The extraordinary nature of this run lies in the sheer velocity of these transitions, which were repeatedly documented on timescales as short as 2.0 minutes. For instance, between JD 2461267.9229 and JD 2461267.9375, the star flickered symmetrically between magnitudes 15.2 and 14.9, completing two full cycles and holding the minimum a third time in a span of just 21 minutes. Because simultaneous checks of the magnitude 15.0 comparison star confirmed it remained completely sharp and steady throughout these movements, these rapid changes cannot be attributed to atmospheric seeing variations.

This observational profile indicates a highly dynamic light curve structure right at the minimum floor. Repetitive breakthroughs of light up to magnitude 14.9 document a succession of rapid, short-timescale brightness fluctuations, causing the system to vary significantly on minute-by-minute timescales before finally closing the session at magnitude 15.2.

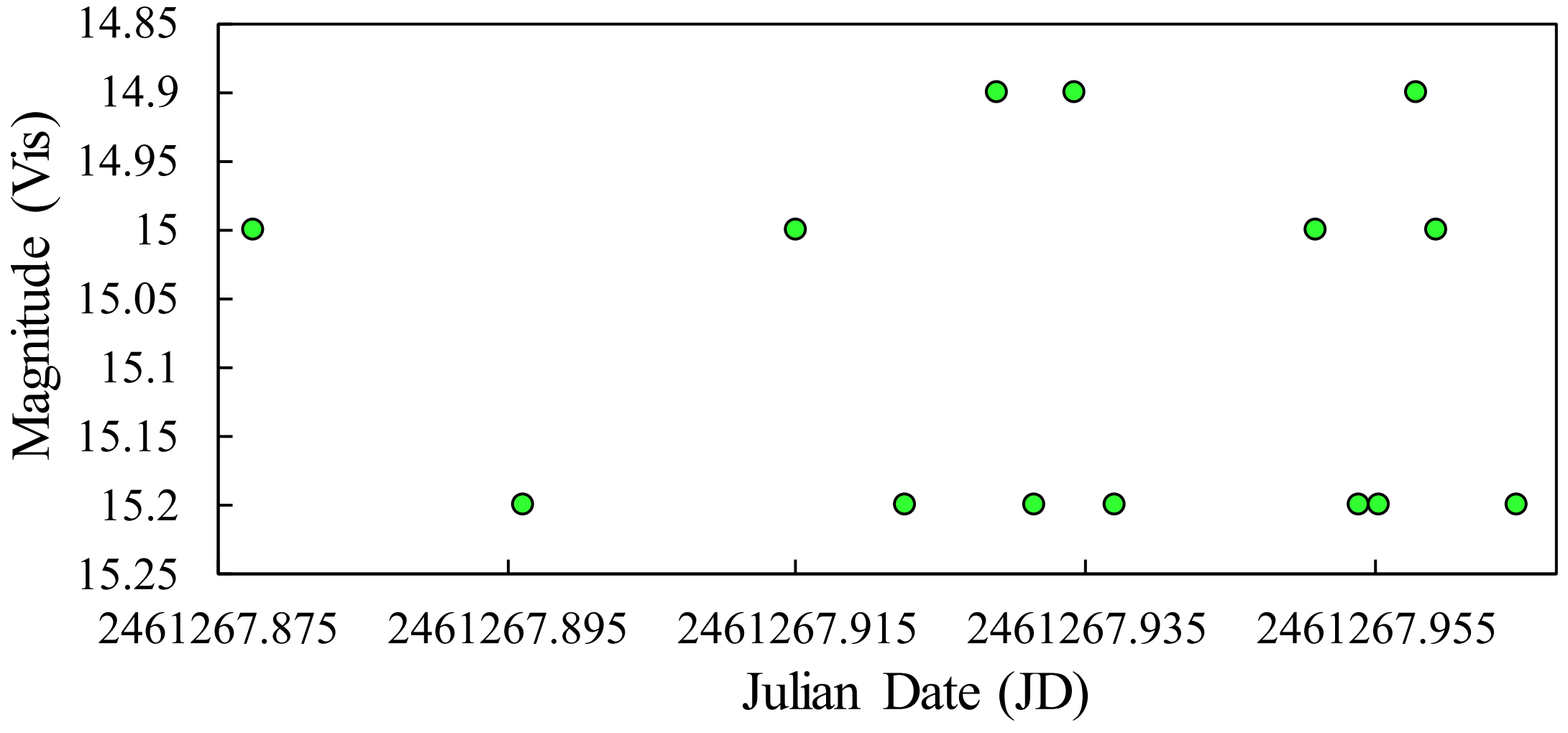

**Figure 12.** High-density unblended light curve for Night 1 (JD 2461267) over an intensive, continuous 2.1-hour eyepiece session. The plot illustrates rapid micro-variability right at the deep minimum, with the star bouncing back and forth between a magnitude 15.2 floor and brief brightening peaks of magnitude 15.0 and 14.9 on timescales as short as 2.0 minutes. Green circles indicate individual visual magnitude estimates.

### 4.1.2 Night 2 (JD 2461268): Structured Stepwise Evolution

**Table 8.** High-density intra-night observational log for Night 2 (JD 2461268)

| **Observed Event** | **Julian Date (JD)** | **Visual Mag** | **Interval (Days)** |
|---|---|---|---|
| Intra-Night Floor Hold | 2461268.8965 | 15.2 | *Session Open* |
| Sustained Floor Observation | 2461268.8986 | 15.2 | 0.0021 |
| Pre-Brightening Base | 2461268.8993 | 15.2 | 0.0007 |
| Initial Brightening Step | 2461268.9007 | 15.1 | 0.0014 |
| Brightening Plateau | 2461268.9014 | 15.1 | 0.0007 |
| Secondary Brightening Shift | 2461268.9028 | 15.0 | 0.0014 |

| Observed Event | Julian Date (JD) | Visual Mag | Interval (Days) |
|---|---|---|---|
| Peak Brightening | 2461268.9035 | 15.0 | 0.0007 |
| Sustained Brightening Peak | 2461268.9049 | 15.0 | 0.0014 |
| Extended Stability Shelf | 2461268.9056 | 15.1 | 0.0007 |
| Shelf Observation Point 2 | 2461268.9063 | 15.1 | 0.0007 |
| Shelf Observation Point 3 | 2461268.9069 | 15.1 | 0.0006 |
| Shelf Observation Point 4 | 2461268.9083 | 15.1 | 0.0014 |
| Shelf Observation Point 5 | 2461268.9097 | 15.1 | 0.0014 |
| Shelf Observation Point 6 | 2461268.9104 | 15.1 | 0.0007 |
| Plateau Shelf / Session Close | 2461268.9111 | 15.1 | 0.0007 |

On the second night, the chaotic variations recorded previously settled into a smooth, structured, symmetric wave spanning a concise 21.0-minute monitoring window. As documented in Table 8 and Figure 13, the session opened at JD 2461268.8965 with a stable floor retention at magnitude 15.2. In a series of rapid 1- to 2-minute steps, the star executed an initial step to magnitude 15.1, climbed to a peak brightening plateau of magnitude 15.0 for roughly three minutes, then settled onto a flat, extended stability shelf at magnitude 15.1, where the star held completely steady until the session closed on JD 2461268.9111.

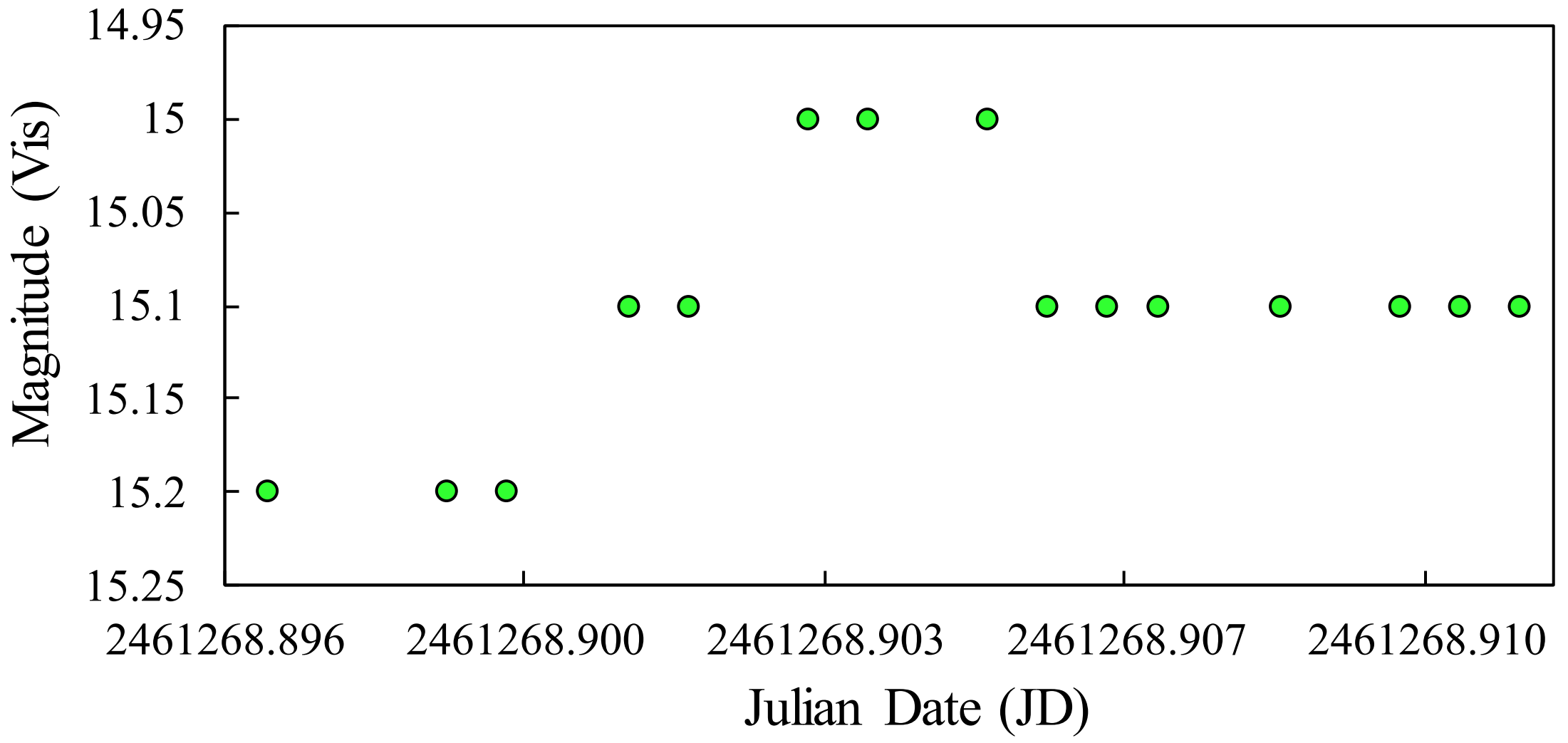

**Figure 13.** High-density unblended light curve for Night 2 (JD 2461268). The profile documents a smooth, step-by-step upward climb away from minimum. The star breaks away from the steady magnitude 15.2 minimum to execute a clear brightening step to magnitude 15.1 and a secondary shift to magnitude 15.0, before settling into a flat, steady shelf pinned at magnitude 15.1 on regular 1-minute logging intervals. Green circles indicate individual visual magnitude estimates.

### 4.1.3 Night 3 (JD 2461269): Rapid Sawtooth Fluctuations

**Table 9.** High-density intra-night observational log for Night 3 (JD 2461269)

| Observed Event | Julian Date (JD) | Visual Mag | Interval (Days) |
|---|---|---|---|
| Stable Opening Plateau | 2461269.9097 | 15.0 | Session Open |
| Plateau Check | 2461269.9104 | 15.0 | 0.0007 |
| Plateau Check | 2461269.9111 | 15.0 | 0.0007 |
| Plateau Exit Step | 2461269.9118 | 15.0 | 0.0007 |
| Minimum Floor | 2461269.9126 | 15.1 | 0.0008 |
| Brightening Peak | 2461269.9132 | 15.0 | 0.0006 |
| Minimum Floor | 2461269.9139 | 15.1 | 0.0007 |
| Continuous Minimum | 2461269.9146 | 15.1 | 0.0007 |
| Brightening Peak | 2461269.9153 | 15.0 | 0.0007 |
| Minimum Floor | 2461269.9160 | 15.1 | 0.0007 |
| Brightening Peak | 2461269.9167 | 15.0 | 0.0007 |
| Minimum Floor | 2461269.9174 | 15.1 | 0.0007 |
| Brightening Peak | 2461269.9181 | 15.0 | 0.0007 |
| Minimum Floor | 2461269.9188 | 15.1 | 0.0007 |
| Brightening Peak | 2461269.9194 | 15.0 | 0.0006 |
| Minimum Floor | 2461269.9201 | 15.1 | 0.0007 |
| Brightening Peak | 2461269.9208 | 15.0 | 0.0007 |
| Minimum Floor | 2461269.9215 | 15.1 | 0.0007 |
| Brightening Peak / Session Close | 2461269.9222 | 15.0 | 0.0007 |

The run on JD 2461269.9097 began with a flat, stable opening plateau near magnitude 15.0. As documented in Table 9 and Figure 14, the stability showed initial movement at JD 2461269.9118 before a distinct micro-variability onset that dropped the star to magnitude 15.1, prior to executing a brief recovery back to magnitude 15.0.

Following JD 2461269.9146, the system initiated a highly regular, alternating sawtooth phase. The star spent the remainder of the session shifting rapidly back and forth between brief intra-night clearing peaks at magnitude 15.0 and a minimum light sequence anchored at magnitude 15.1 on a tight, systematic 1-minute interval. This observed profile demonstrates that the light curve profile remained highly unsettled before closing the session with a final peak observation at magnitude 15.0 at JD 2461269.9222.

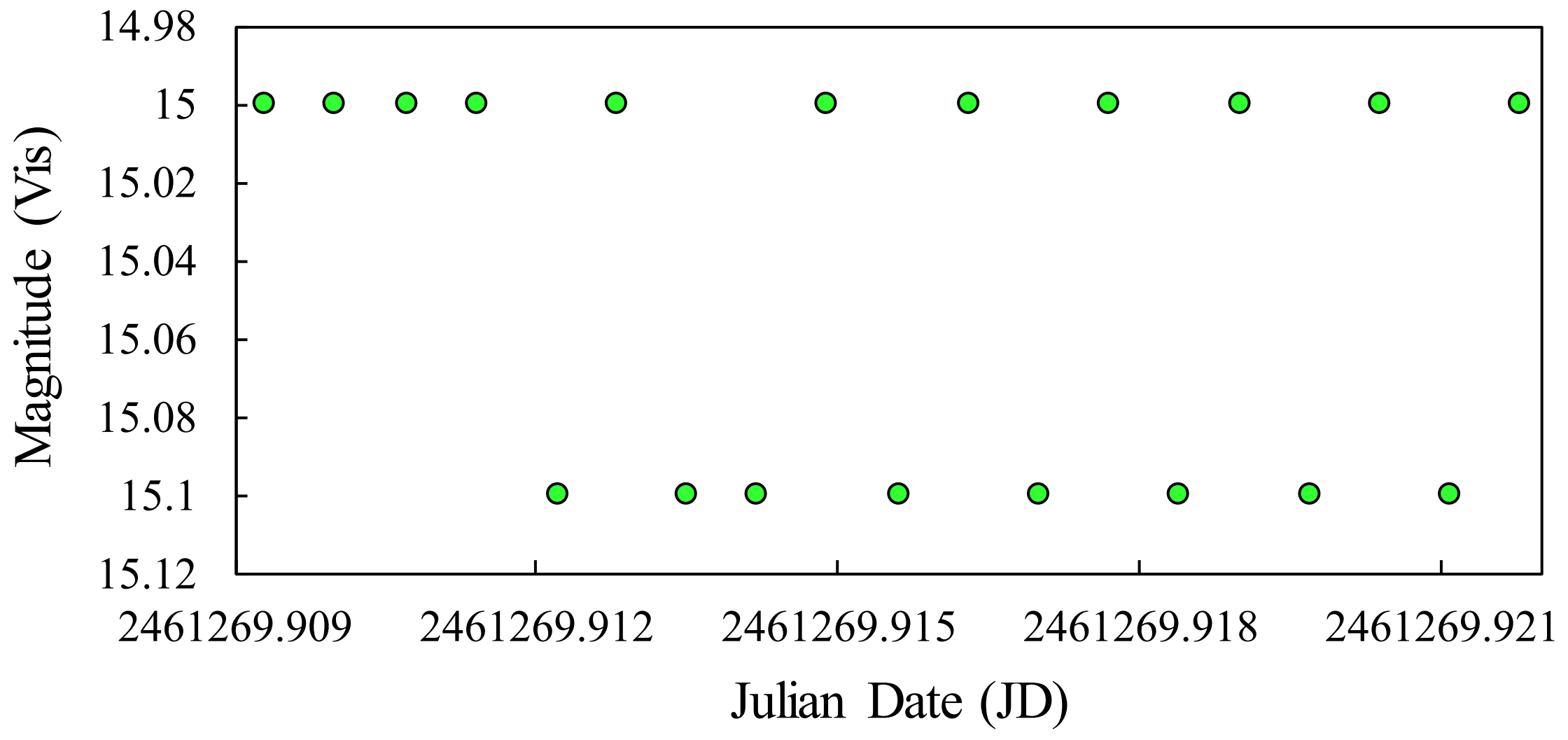

**Figure 14.** High-density unblended light curve for Night 3 (JD 2461269). The dataset demonstrates a rapid sawtooth variability pattern. Following a brief opening plateau at magnitude 15.0, the star enters a sudden micro-variability onset shifting rapidly back and forth between brief magnitude 15.0 peaks and a continuous magnitude 15.1 base. Green circles indicate individual visual magnitude estimates.

#### 4.1.4 Night 4 (JD 2461270): The Sub-Minute Micro-Variability Horizon

**Table 10.** High-density intra-night observational log for Night 4 (JD 2461270)

| Observed Event | Julian Date (JD) | Visual Mag | Interval (Days) |
|---|---|---|---|
| Micro-Variability Phase Onset | 2461270.8750 | 15.1 | *Session Open* |
| Sub-Minute Micro-Fluctuation | 2461270.8757 | 15.1 | 0.0007 |
| Sustained Micro-Fluctuation 3 | 2461270.8764 | 15.1 | 0.0007 |
| Sustained Micro-Fluctuation 4 | 2461270.8771 | 15.1 | 0.0007 |
| Sustained Micro-Fluctuation 5 | 2461270.8778 | 15.1 | 0.0007 |
| Sustained Micro-Fluctuation 6 | 2461270.8785 | 15.1 | 0.0007 |
| Sustained Micro-Fluctuation 7 | 2461270.8792 | 15.1 | 0.0007 |
| Sustained Micro-Fluctuation 8 | 2461270.8799 | 15.1 | 0.0007 |
| Sustained Micro-Fluctuation 9 | 2461270.8806 | 15.1 | 0.0007 |
| Session Close | 2461270.8813 | 15.1 | 0.0007 |

The final night of the intensive sequence showed a distinct change in how the star behaved. The star opened at magnitude 15.1 on JD 2461270.8750. As documented in Table 10 and shown in Figure 15, the amplitude swings seen on the previous nights completely disappeared. Instead, the light curve flattened out, holding steady at magnitude 15.1 over a 9-minute interval. However, looking closely at 270× magnification against the steady magnitude 15.0 comparison star, the target was not perfectly still. It was constantly exhibiting rapid, low-amplitude light variations in between the one-minute entries. These fast movements happened too rapidly to break down and record as separate magnitude steps by eye. Even though the logged points show a flat line at magnitude 15.1, the star was clearly highly unsettled, with its apparent light fluctuating rapidly on a second-by-second basis right in front of the eyepiece.

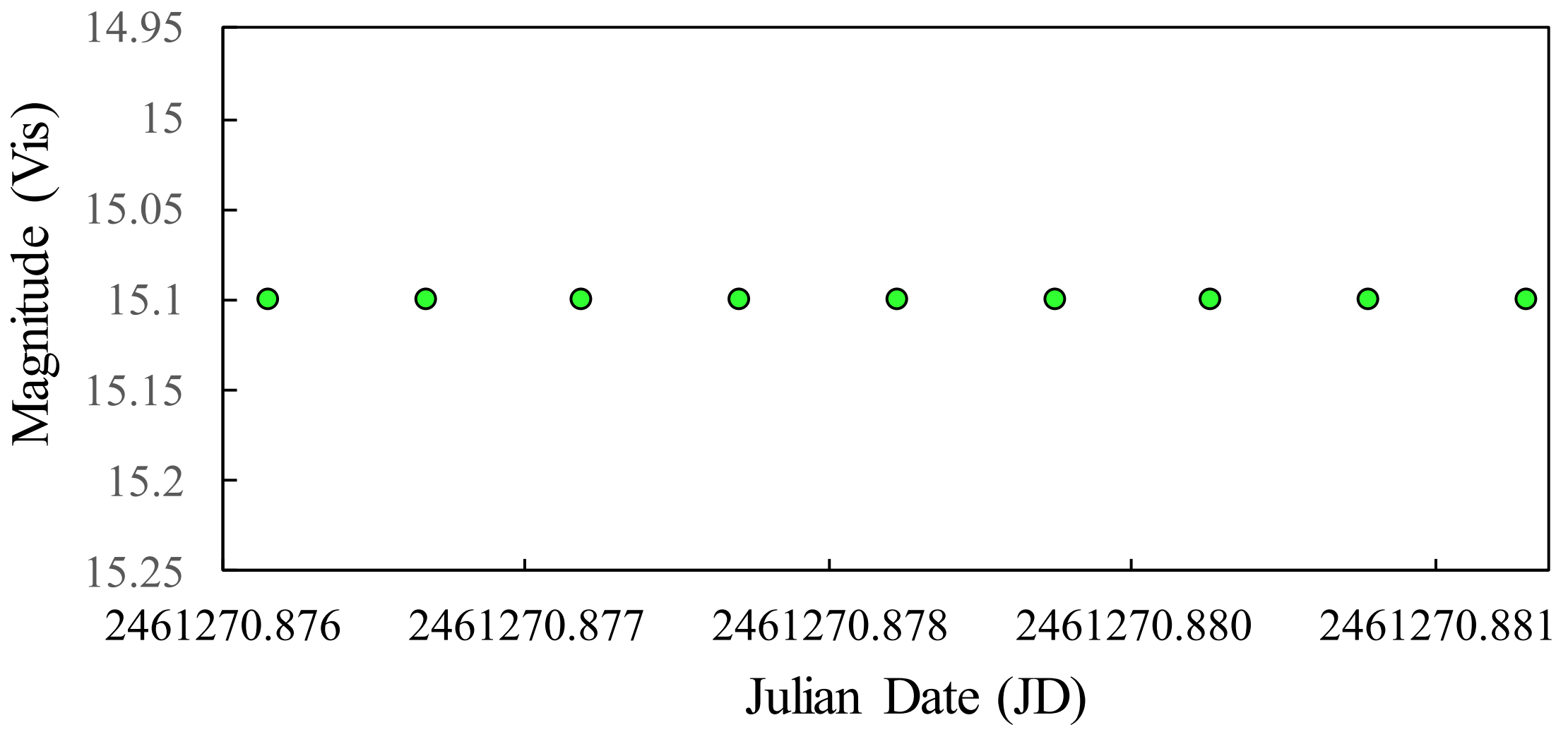

**Figure 15.** High-density unblended light curve for Night 4 (JD 2461270). The plot displays a complete flattening of macroscopic movements, with the star holding a visually uniform sequence locked at magnitude 15.1 across the entire session. High-magnification comparisons against the steady magnitude 15.0 comparison star simultaneously confirmed a persistent, sub-minute shimmering behaviour operating entirely beneath the logging limits of my eye. Green circles indicate individual visual magnitude estimates.

### 4.2 Late-Epoch Monitoring Campaign (Nights 5 to 12: JD 2461281–2461321)

To evaluate the long-term structural persistence of the heavy stellar obscuration at its deepest fading limit, a second intensive high-density campaign was initiated eleven days later following the close of the initial observing phase. Because the visual profile variations quickly proved to be highly consistent across these subsequent sessions, these observations were initially conducted as shorter-duration runs, but extended at the end because of the rapid variations.

#### 4.2.1 Night 5 (JD 2461281): Recurrent High-Frequency Flickering

**Table 11.** High-frequency intra-night observational log for Night 5 (JD 2461281)

| Observed Event | Julian Date (JD) | Visual Mag | Interval (Days) |
|---|---|---|---|
| Late-Epoch Campaign Opening | 2461281.8764 | 15.2 | *Session Open* |
| Initial Brightening Shift | 2461281.8778 | 15.1 | 0.0014 |
| Brightening Plateau | 2461281.8792 | 15.1 | 0.0014 |
| Rapid Minimum Reversion | 2461281.8799 | 15.2 | 0.0007 |
| Transient Brightening Peak | 2461281.8813 | 15.0 | 0.0014 |
| Minimum Floor Reversion / Session Close | 2461281.8819 | 15.2 | 0.0006 |

The opening session of the late-epoch campaign, spanning a concise 7.9-minute observing window, confirmed that the light curve profile remained in a highly dynamic and unsettled state. As documented in Table 11 and shown in Figure 16, the star opened the run locked flat at its absolute minimum floor of magnitude 15.2 before immediately executing a 0.1-magnitude shift to a brief plateau at magnitude 15.1. The most remarkable feature of this short run occurred at JD 2461281.8813, where a rapid, high-amplitude clearing drove the star up to magnitude 15.0, a full 0.2-magnitude brightness jump achieved in a span of just 2.0 minutes. This peak was short-lived, as the system collapsed completely back to its absolute floor of magnitude 15.2 less than one minute later on JD 2461281.8819.

The observational profile proves that the highly broken-up nature of the fading profile was not a temporary anomaly of the JD 2461267 epoch. Instead, the persistent, fast-paced transitions between magnitude 15.0 and 15.2 reveal that the light curve structure remained dominated by the same high-density variations eleven days later.

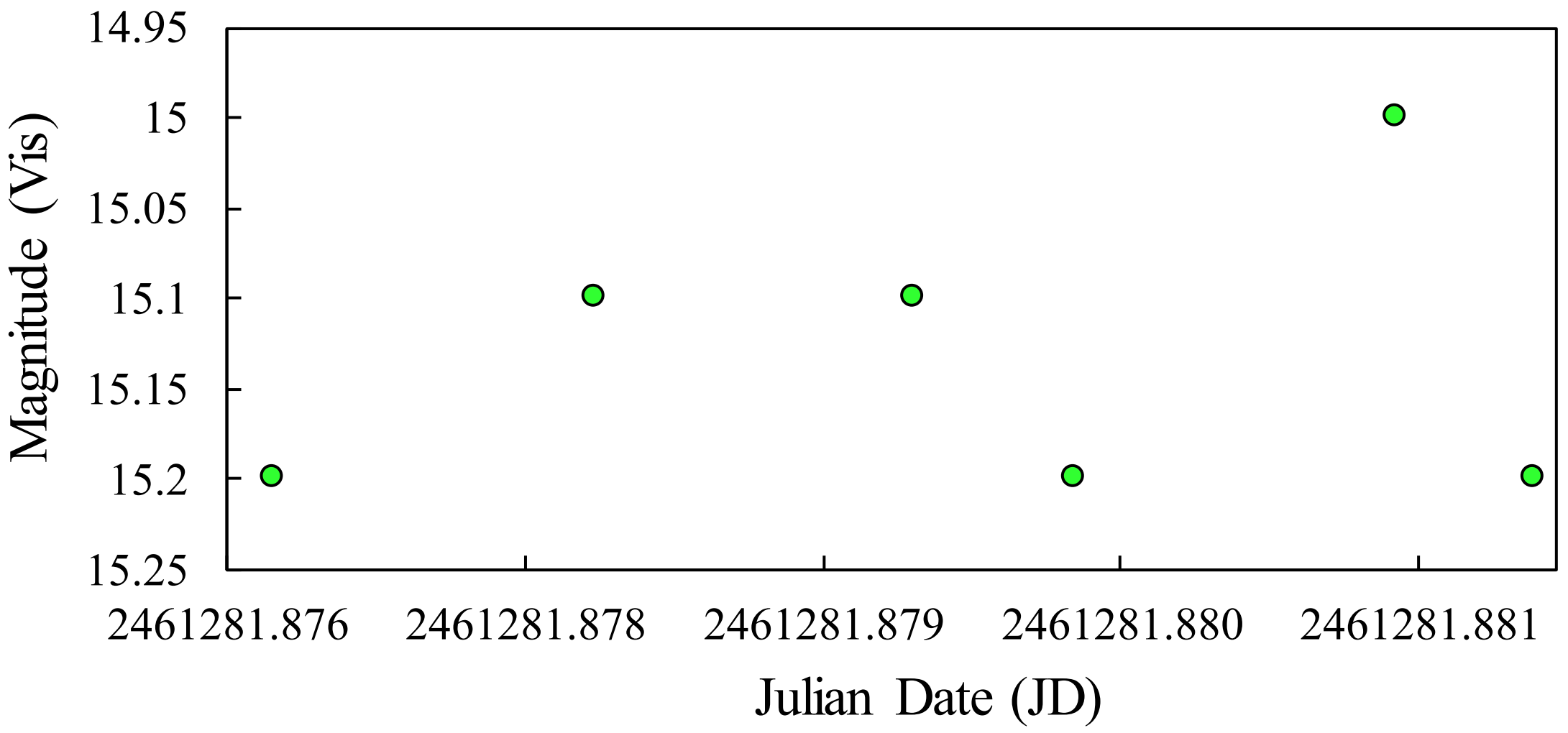

**Figure 16.** High-density unblended light curve for the opening session of the late-epoch campaign on Night 5 (JD 2461281). The profile displays recurrent high-frequency flickering at the deep limit, highlighted by an immediate brightening step from a magnitude 15.2 base to a 15.1 plateau, followed by an exceptionally fast 0.2-magnitude clearing peak up to magnitude 15.0 achieved over a span of just 2.0 minutes. Green circles indicate individual visual magnitude estimates.

### 4.2.2 Night 6 (JD 2461282): Alternating Sub-Minute Flickering

**Table 12.** High-density intra-night observational log for Night 6 (JD 2461282)

| Observed Event | Julian Date (JD) | Visual Mag | Interval (Days) |
|---|---|---|---|
| Late-Epoch Session Opening | 2461282.8813 | 15.1 | *Session Open* |
| Rapid Minimum Reversion | 2461282.8819 | 15.2 | 0.0006 |
| Transient Brightening Shift | 2461282.8826 | 15.1 | 0.0007 |
| Minimum Floor Reversion / Session Close | 2461282.8833 | 15.2 | 0.0007 |

The second session of the late-epoch campaign, spanning a concise 2.9-minute monitoring window, recorded a rapid symmetrical sawtooth pattern. As documented in Table 12 and shown in Figure 17, the session opened at magnitude 15.1 before abruptly plunging to the absolute floor of magnitude 15.2 in less than a minute. The star immediately reversed direction, executing a transient brightening shift back to magnitude 15.1 over a 1-minute interval, before returning cleanly to the 15.2 minimum floor at the session close.

This alternating sequence mimics the rapid sawtooth variations documented on Night 3, but on an accelerated timescale. The clean, rapid bouncing between 15.1 and 15.2 shows that the light curve profile remained highly irregular at 1-minute intervals.

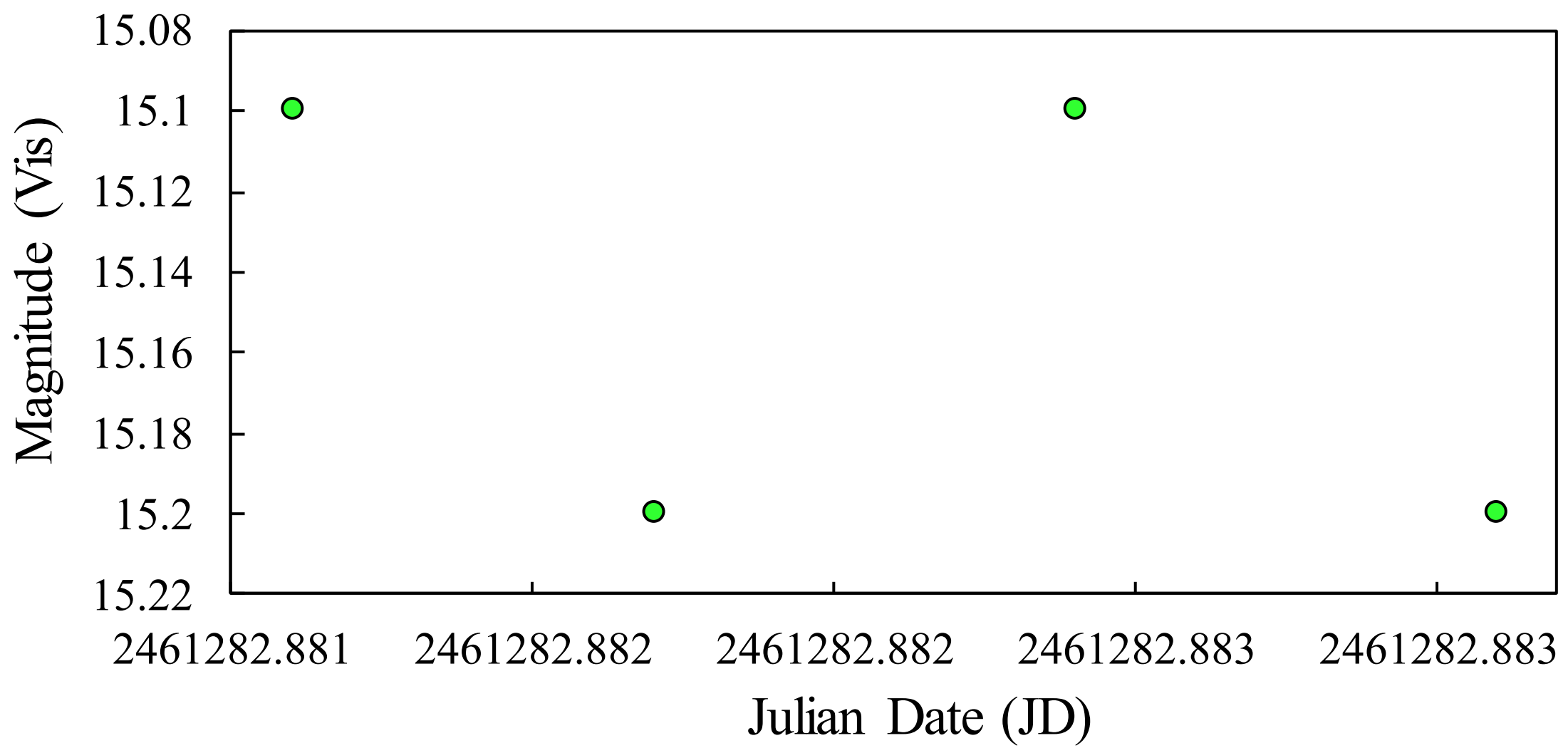

**Figure 17.** High-density unblended light curve for Night 6 (JD 2461282). This plot illustrates an alternating sub-minute flickering sequence. The star executes rapid, symmetrical sawtooth reversals, bouncing cleanly back and forth between magnitudes 15.1 and 15.2 in a structured alternating sequence on 1-minute timescales. Green circles indicate individual visual magnitude estimates.

### 4.2.3 Night 7 (JD 2461283): Rhythmic Step and Oscillation

**Table 13.** High-density observational log of intra-night oscillations for Night 7 (JD 2461283)

| Observed Event | Julian Date (JD) | Visual Mag | Interval (Days) |
|---|---|---|---|
| Late-Epoch Session Opening | 2461283.8868 | 15.2 | *Session Open* |
| Initial Brightening Step | 2461283.8875 | 15.1 | 0.0007 |
| Brightening Plateau Hold | 2461283.8882 | 15.1 | 0.0007 |
| Minimum Floor Reversion | 2461283.8889 | 15.2 | 0.0007 |
| Secondary Brightening Step | 2461283.8896 | 15.1 | 0.0007 |
| Minimum Floor Reversion / Session Close | 2461283.8903 | 15.2 | 0.0007 |

The third session of the late-epoch campaign, documented in Table 13 and shown in Figure 18 and spanning a precise 5.0-minute monitoring interval, revealed a mixture of plateau stability and rapid, structured fluctuations. The run opened with the star at its deep floor of magnitude 15.2 before it immediately climbed to a brightening shelf at magnitude 15.1. The star managed to hold this elevated state flat for a two-minute window before the stability broke.

Immediately following JD 2461283.8882, the system resumed a rapid, high-frequency sawtooth pattern, bouncing cleanly between magnitudes 15.2 and 15.1 on an unvarying 1-minute interval before closing at its absolute minimum threshold. This session is highly significant because it perfectly replicates the exact 1-minute timeline structure seen on Night 6. This reproducibility provides direct observational evidence that the physical magnitude variations are not random chaotic noise, but instead reveal a highly organised, systematic pattern within the star's short-timescale light curve profile.

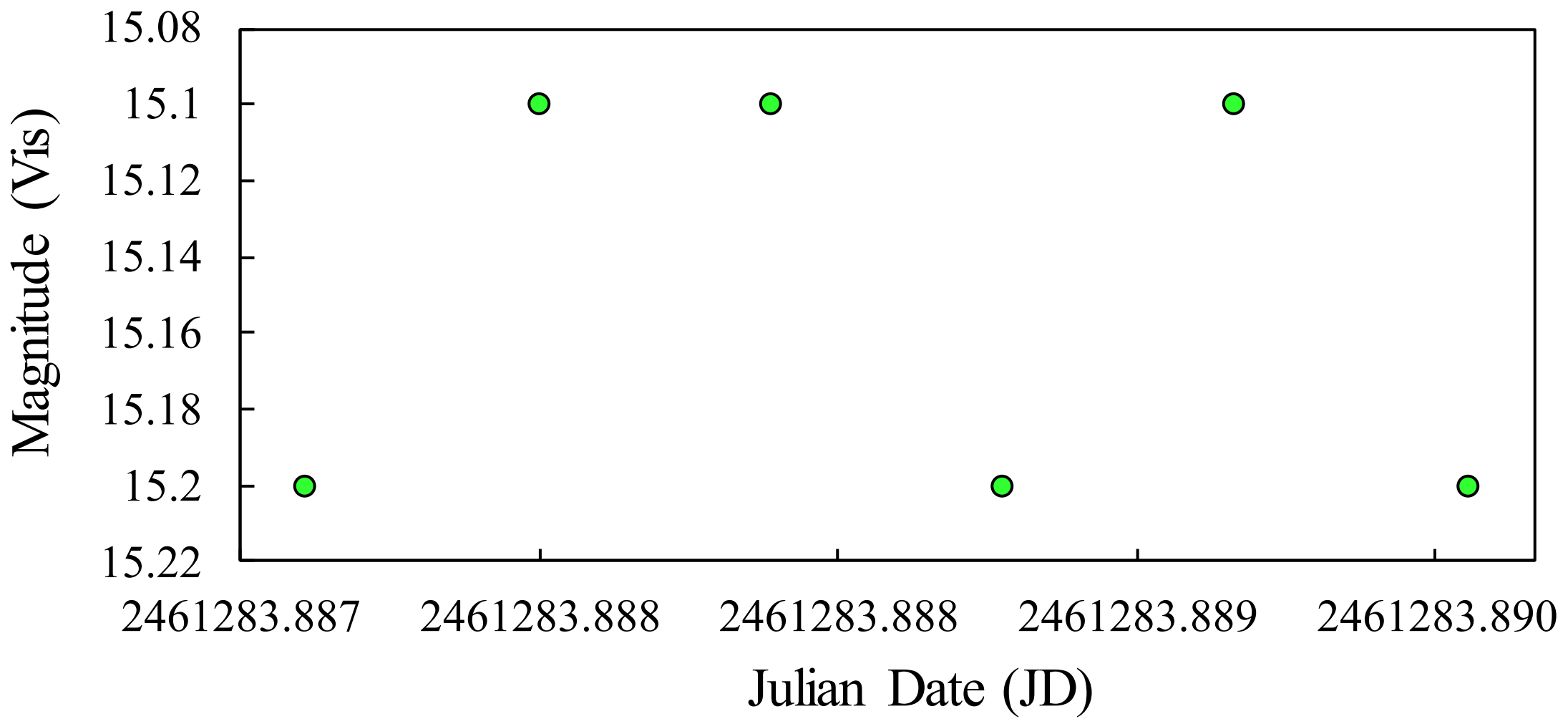

**Figure 18.** High-density light curve for Night 7 (JD 2461283). The plot illustrates a distinct combination of plateau stability and rapid oscillations. The star opens with an immediate brightening step from the floor of magnitude 15.2, to a brightening plateau at magnitude 15.1, holding steady for exactly two minutes before breaking downward into a structured, 1-minute sawtooth sequence between magnitudes 15.1 and 15.2. Green circles indicate individual visual magnitude estimates.

### 4.2.4 Night 8 (JD 2461285): Extended Alternating Zigzag Run

**Table 14.** High-density intra-night observational log for Night 8 (JD 2461285)

| Observed Event | Julian Date (JD) | Visual Mag | Interval (Days) |
|---|---|---|---|
| Late-Epoch Session Opening | 2461285.8819 | 15.2 | *Session Open* |
| Initial Brightening Step | 2461285.8826 | 15.1 | 0.0007 |
| Brightening Plateau Hold | 2461285.8833 | 15.1 | 0.0007 |
| Minimum Floor Reversion | 2461285.8840 | 15.2 | 0.0007 |
| Structured Brightening Shift | 2461285.8847 | 15.1 | 0.0007 |
| Minimum Floor Reversion | 2461285.8854 | 15.2 | 0.0007 |
| Secondary Brightening Shift | 2461285.8861 | 15.1 | 0.0007 |
| Minimum Floor Reversion | 2461285.8868 | 15.2 | 0.0007 |
| Brightening Step / Session Close | 2461285.8875 | 15.1 | 0.0007 |

The fourth session of the late-epoch campaign, documented in Table 14 and shown in Figure 19 and spanning a precise 8.1-minute monitoring interval, documented a highly sustained sequence of rapid variations. The run opened with the star at its deep floor of magnitude 15.2 on JD 2461285.8819, before it immediately moved up to an elevated plateau at magnitude 15.1, holding steady for a 1-minute interval. Following JD 2461285.8833, the star broke out of its brief hold and initiated a long unbroken chain of rapid 1-minute sawtooth oscillations. The target fluctuated cleanly back and forth between magnitudes 15.1 and 15.2 in a highly structured, alternating sequence, repeating this cycle before closing the session on a bright clearing peak at magnitude 15.1.

The long duration of these steady 1-minute brightness reversals shows just how turbulent the system remained. Rather than a temporary fluctuation, this run proves that the light curve profile remained characterised by high-density, periodic variations, causing regular amplitude changes operating on a minute-by-minute sequence.

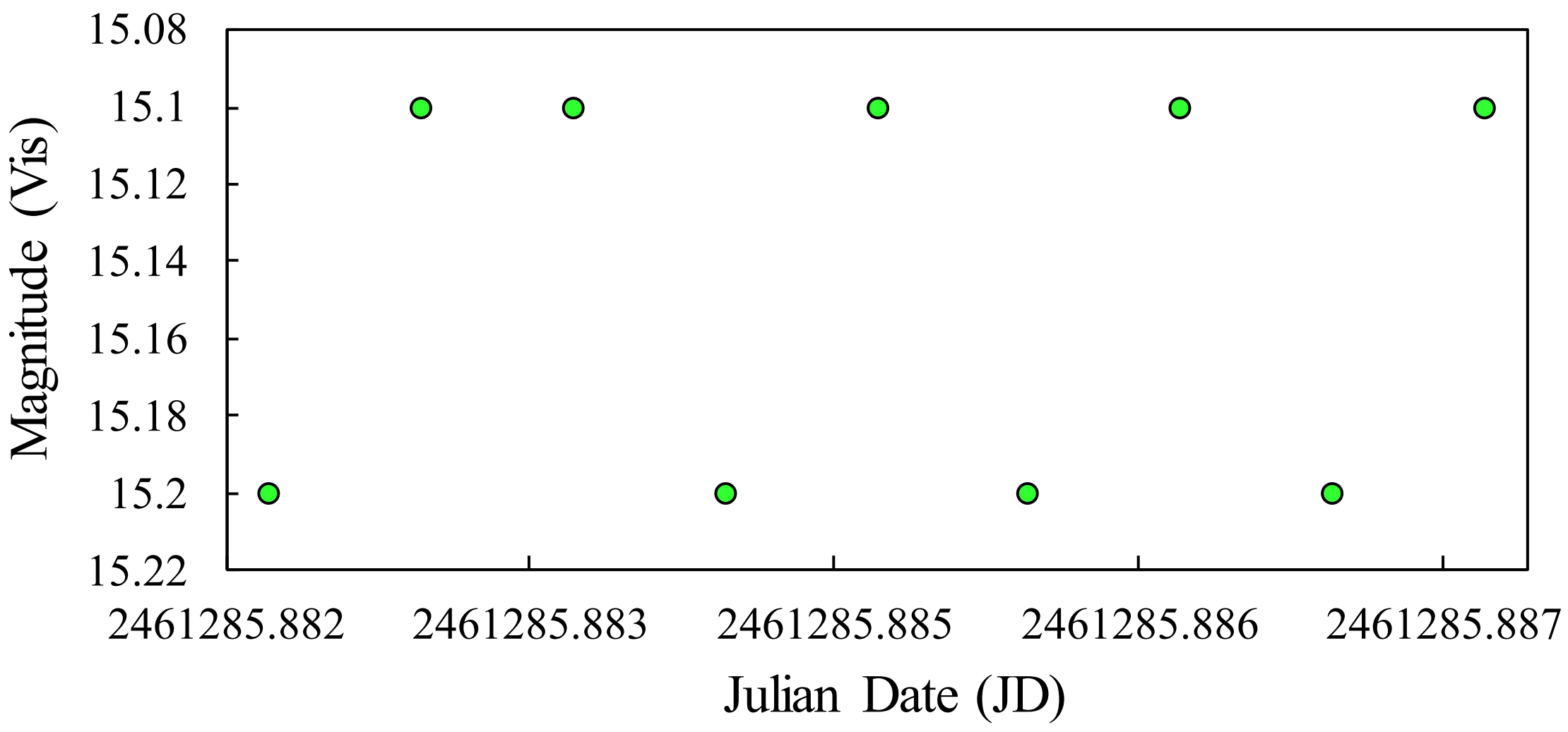

**Figure 19.** High-density unblended light curve for Night 8 (JD 2461285). The dataset documents a highly sustained, extensive run of rapid variations. Following an initial brightening step and a brief two-minute brightening plateau hold at magnitude 15.1, the star enters a long, unbroken chain of consecutive 1-minute alternating sawtooth cycles bouncing cleanly between magnitudes 15.1 and 15.2. Green circles indicate individual visual magnitude estimates.

### 4.2.5 Night 9 (JD 2461286): Complex Oscillating Plateau

**Table 15.** High-density intra-night observational log for Night 9 (JD 2461286)

| Observed Event | Julian Date (JD) | Visual Mag | Interval (Days) |
|---|---|---|---|
| Late-Epoch Campaign Opening | 2461286.8896 | 15.2 | *Session Open* |
| Initial Brightening Step | 2461286.8903 | 15.1 | 0.0007 |
| Structured Minimum Reversion | 2461286.8910 | 15.2 | 0.0007 |
| Secondary Brightening Shift | 2461286.8917 | 15.1 | 0.0007 |
| Brightening Plateau Onset | 2461286.8924 | 15.1 | 0.0007 |
| Rapid Minimum Reversion | 2461286.8931 | 15.2 | 0.0007 |
| Third Brightening Shift | 2461286.8938 | 15.1 | 0.0007 |
| Extended Plateau Onset | 2461286.8944 | 15.1 | 0.0006 |
| Minimum Floor Reversion | 2461286.8951 | 15.2 | 0.0007 |
| Fourth Brightening Shift | 2461286.8958 | 15.1 | 0.0007 |
| Late-Session Plateau Hold | 2461286.8965 | 15.1 | 0.0007 |
| Minimum Floor / Session Close | 2461286.8972 | 15.2 | 0.0007 |

The fifth session of the late-epoch campaign, documented in Table 15 and shown in Figure 20 and spanning an intensive 10.9-minute monitoring window, recorded a highly complex combination of rapid oscillations and brief plateau stability. The run opened with the star at its deep floor of magnitude 15.2 on JD 2461286.8896, before executing an initial rapid 1-minute sawtooth cycle that bounced sharply between magnitudes 15.1 and 15.2.

Following JD 2461286.8917, the behaviour shifted into a structured sequence of short-lived plateaus. The star held a flat plateau at magnitude 15.1 across two consecutive 1-minute entries before reverting to the floor, immediately followed by another multi-point brightening step that formed a second, two-point plateau encompassing the onset at JD 2461286.8944. A final, rapid sawtooth oscillation carried the star through a fourth brightness peak at magnitude 15.1 on JD 2461286.8965, before the target permanently lost momentum, dropping flat into a deep state at magnitude 15.2 on JD 2461286.8972 to close the session.

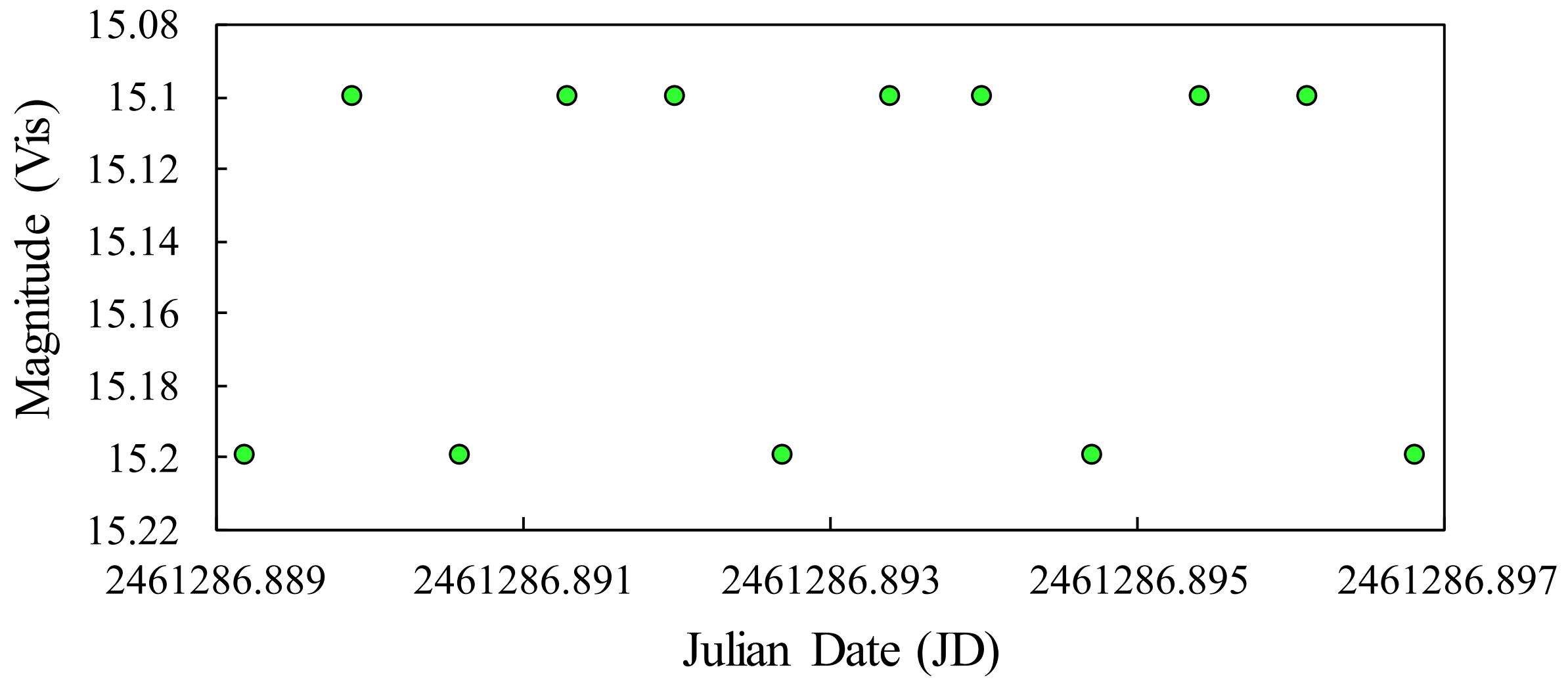

**Figure 20.** High-density unblended light curve for Night 9 (JD 2461286). The plot documents a complex combination of rapid oscillations and short-lived plateau steps. The star sweeps through two initial 1-minute sawtooth cycles before stabilising along consecutive short plateaus at magnitude 15.1, broken by sudden floor drops, before locking into its final minimum at magnitude 15.2. Green circles indicate individual visual magnitude estimates.

### 4.2.6 Night 10 (JD 2461290): Deep Ingress and Intermediate Plateau Oscillations

**Table 16.** High-density intra-night observational log for Night 10 (JD 2461290)

| Observed Event | Julian Date (JD) | Visual Mag | Interval (Days) |
|---|---|---|---|
| Late-Epoch Campaign Session Opening | 2461290.9076 | 15.1 | *Session Open* |
| Rapid Minimum Reversion | 2461290.9083 | 15.2 | 0.0007 |
| Intermediate Plateau Recovery | 2461290.9090 | 15.15 | 0.0007 |
| Primary Brightening Shift | 2461290.9097 | 15.1 | 0.0007 |
| Intermediate Plateau Reversion | 2461290.9104 | 15.15 | 0.0007 |
| Sustained Intermediate Plateau Point 2 | 2461290.9111 | 15.15 | 0.0007 |
| Sustained Intermediate Plateau Point 3 | 2461290.9118 | 15.15 | 0.0007 |
| Secondary Brightening Shift | 2461290.9125 | 15.1 | 0.0007 |
| Intermediate Plateau Reversion Point 4 | 2461290.9132 | 15.15 | 0.0007 |
| Third Brightening Shift | 2461290.9139 | 15.1 | 0.0007 |
| Intermediate Plateau Reversion Point 5 | 2461290.9146 | 15.15 | 0.0007 |
| Sustained Intermediate Plateau Point 6 | 2461290.9153 | 15.15 | 0.0007 |
| Sustained Intermediate Plateau Point 7 | 2461290.9160 | 15.15 | 0.0007 |
| Sustained Intermediate Plateau Point 8 | 2461290.9167 | 15.15 | 0.0007 |
| Final Brightening Peak / Session Close | 2461290.9174 | 15.1 | 0.0007 |

The tenth high-density session presented in Table 16 and shown in Figure 21, documented a fascinating change in the light profile behaviour, displaying a highly structured intermediate plateau sequence. The session opened at magnitude 15.1 on JD 2461290.9076, before experiencing a rapid, deep downward plunge to magnitude 15.2 over the 1-minute step. The star immediately rebounded, establishing a prolonged, tight stability plateau at an intermediate threshold of magnitude 15.15. The significance of this run lies in how the star fluctuated against this 15.15 intermediate baseline. Rather than bouncing completely down to magnitude 15.2, the variations shifted into a very tight 0.05-magnitude range, forming a steady plateau sequence at magnitude 15.15 that was broken up by sharp, temporary brightening shifts up to magnitude 15.1 during the 1-minute intervals. At magnitude 15.15, the direct starlight from the stellar core remained heavily attenuated, but the ultra-rapid, 1-minute clearings to magnitude 15.1 confirm that the star was still highly dynamic to the session close.

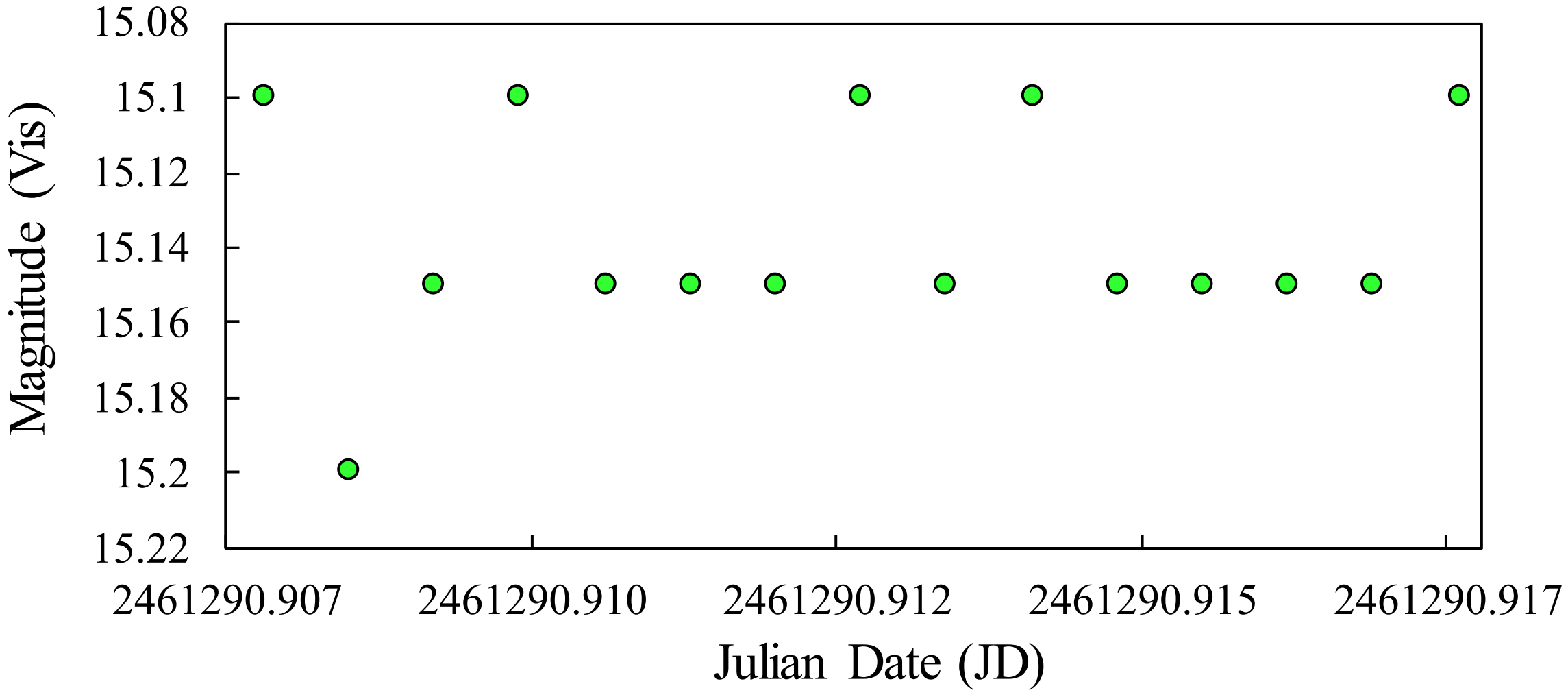

**Figure 21.** High-density unblended light curve for Night 10 (JD 2461290). The plot illustrates an initial rapid plunge, from magnitude 15.1 straight down to an absolute floor at magnitude 15.2 in exactly one minute, before establishing a prolonged intermediate plateau sequence. Rather than full drops to the floor, the subsequent variations shifted into a tight 0.05-magnitude range, holding flat at an intermediate threshold of magnitude 15.15 broken periodically by rapid 1-minute brightening back to magnitude 15.1. Green circles indicate visual estimates.

### 4.2.7 Night 11 (JD 2461318): Rhythmic Alternations at the Absolute Minimum Floor

**Table 17.** High-density intra-night observational log for Night 11 (JD 2461318)

| Observed Event | Julian Date (JD) | Visual Mag | Interval (Days) |
|---|---|---|---|
| Deep Minimum Session Opening | 2461318.9035 | 15.2 | *Session Open* |
| Initial Brightening Shift | 2461318.9042 | 15.15 | 0.0007 |
| Structured Minimum Reversion | 2461318.9049 | 15.2 | 0.0007 |
| Secondary Brightening Shift | 2461318.9056 | 15.15 | 0.0007 |
| Symmetrical Minimum Reversion | 2461318.9063 | 15.2 | 0.0007 |
| Third Brightening Shift | 2461318.9069 | 15.15 | 0.0006 |
| Secondary Symmetrical Reversion | 2461318.9076 | 15.2 | 0.0007 |
| Fourth Brightening Shift | 2461318.9083 | 15.15 | 0.0007 |
| Minimum Floor Reversion / Session Close | 2461318.9090 | 15.2 | 0.0007 |

The eleventh high-density session documented in Table 17 and shown in Figure 22 and spanning an approximate 7.9-minute window, documents a profound qualitative shift in the micro-variability of the system, observing an alternating pattern right at the base of the deep minimum. The star initiated a rapid sawtooth sequence, alternating cleanly between a minimum floor of magnitude 15.2 and a brightening peak of magnitude 15.15 for four consecutive cycles. The significance of this run lies in how the star fluctuated across a tight 0.05-magnitude range. Because the system was caught at its absolute historical minimum, the direct point-source starlight from the stellar core remained heavily obscured. The rapid, highly structured 1-minute clearings up to magnitude 15.15 reveal that even at this deep limit, the starlight exhibits periodic variations on a minute-by-minute sequence.

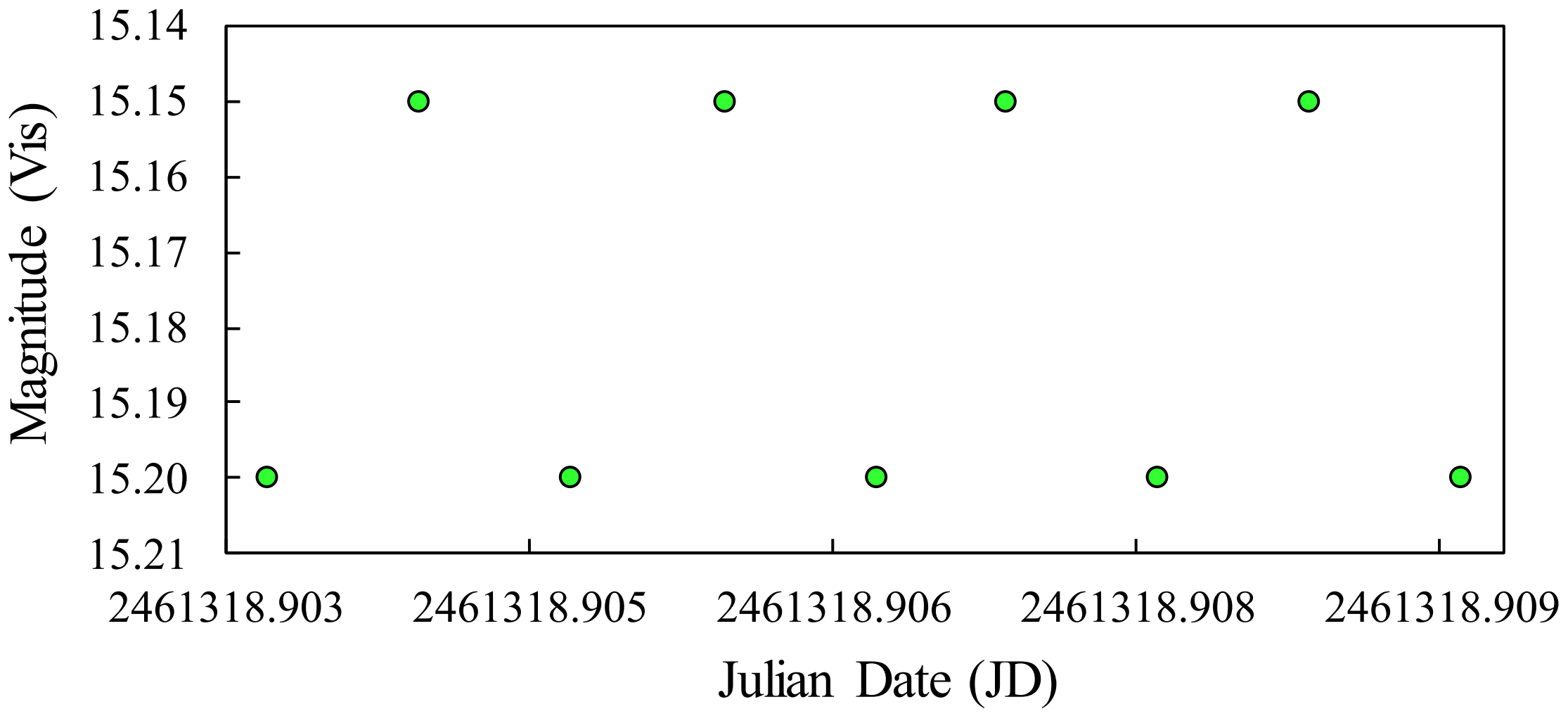

**Figure 22.** High-density unblended light curve highlighting the deep minimum on Night 11 (JD 2461318). The plot illustrates a highly regular, alternating bimodal sequence. The variations are constrained to a tight 0.05-magnitude range, with the star bouncing cleanly between magnitudes 15.2 and 15.15 at 1-minute intervals. Green circles indicate individual visual magnitude estimates.

### 4.2.8 Night 12 (JD 2461321): Failed Core Breakouts and Total Floor Lock

**Table 18.** High-density intra-night observational log for Night 12 (JD 2461321)

| Observed Event | Julian Date (JD) | Visual Mag | Interval (Days) |
|---|---|---|---|
| Deep Minimum Session Opening | 2461321.9069 | 15.2 | *Session Open* |
| Sustained Core Obscuration | 2461321.9076 | 15.2 | 0.0007 |
| Ongoing Core Attenuation | 2461321.9083 | 15.2 | 0.0007 |
| Ongoing Core Attenuation Point 2 | 2461321.9090 | 15.2 | 0.0007 |
| Ongoing Core Attenuation Point 3 | 2461321.9097 | 15.2 | 0.0007 |
| Ongoing Core Attenuation Point 4 | 2461321.9104 | 15.2 | 0.0007 |
| Session Close | 2461321.9111 | 15.2 | 0.0007 |

The twelfth high-density session, documented in Table 18 and shown in Figure 23 and spanning an approximate 6.0-minute interval, documents the absolute limits of point-source detection during the deep fading on JD 2461321.9069. Throughout this continuous observation run, the star remained completely incapable of executing the high-frequency magnitude 15.1 or 15.15 clearing variations seen during prior epochs. Instead, the star was noticeably flat at magnitude 15.2 at 1-minute intervals. This continuous run represents a distinct optical profile, where the star was heavily obscured across successive observations, demonstrating an extended phase where the pinpoint core would momentarily soften or harden slightly while remaining completely confined beneath the deep baseline floor. Because the core failed to re-emerge as a bright pinpoint, it remained locked at its minimum floor.

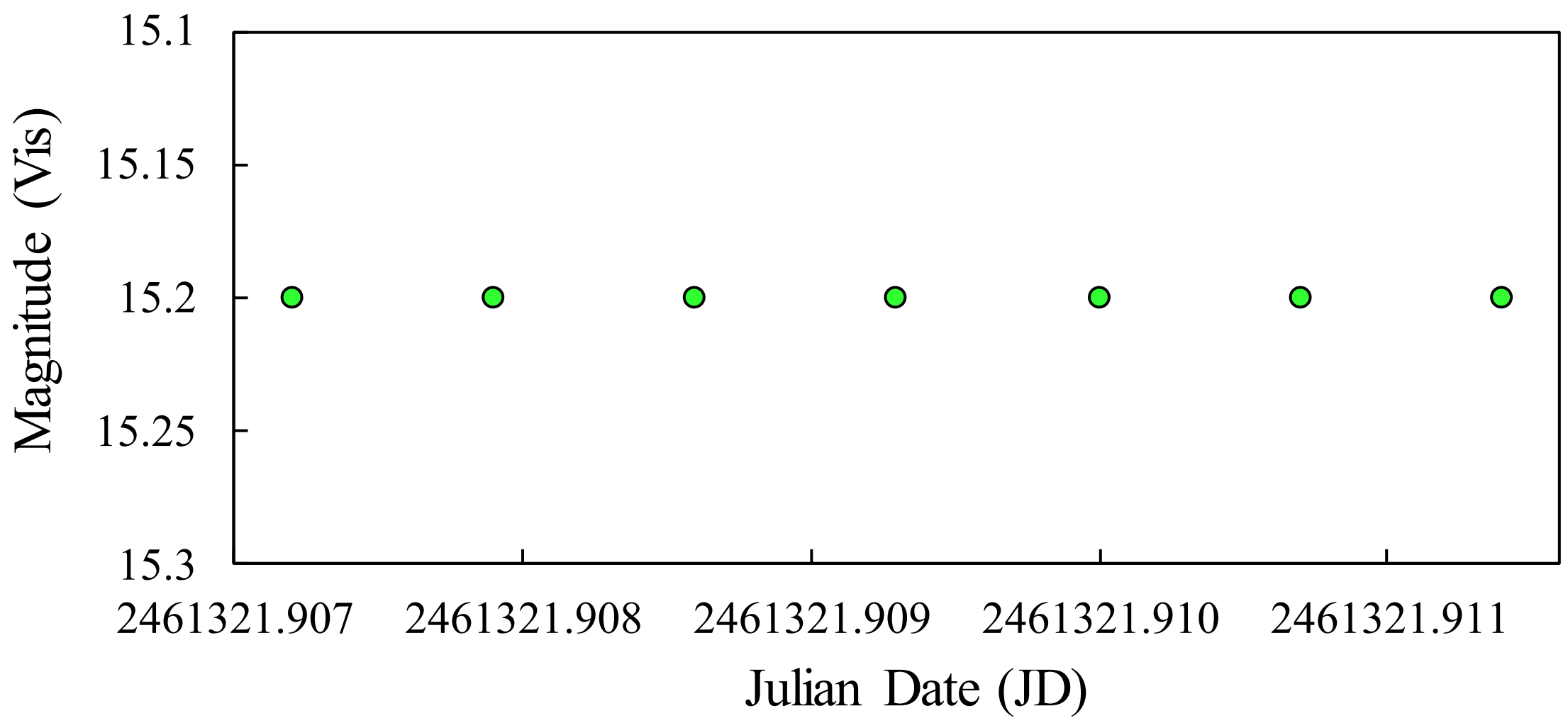

**Figure 23.** High-density unblended light curve for Night 12 (JD 2461321). The plot highlights a complete flattening of variations, with the target pinned flat at magnitude 15.2 due to the continuous failure of the pinpoint stellar core to break through the dust cloud. Green circles indicate individual visual magnitude estimates.

### 4.3 Physical Implications of Sub-Minute Pinpoint Core Variations

The high-density observations compiled across monitoring campaigns provided direct visual proof that the observed variations occur on timescales far shorter than the light curves suggest. The crucial diagnostic marker remains the physical behaviour of the target's point-source core. While the 0.05- to 0.1-magnitude variations logged on 1-minute intervals (Tables 8-18) are distinct, the star simultaneously brightens and fades between these entries on timescales near 30 seconds or less.

This ultra-fast activity is easily detected at the eyepiece because the star's pinpoint stellar core actively fades out into the diffuse halo and immediately reappears against the steady magnitude 15.0 comparison star. Because the comparison star remains consistently sharp throughout these runs, atmospheric seeing is completely ruled out.

At magnitude 15.2, the direct point-source starlight from the core is fully obscured, leaving only the faint halo of scattered circumstellar light visible. The ultra-rapid, 30-second reappearance of the pinpoint stellar core at magnitude 15.15 marks the exact moments direct starlight breaks through, before the stellar point source returns to its heavily attenuated minimum state. During the deepest parts of this minimum, the star exhibited a highly unique behaviour that directly changed how the observations had to be recorded. On many nights, I noted that the stellar point source experienced rapid profile changes. The pinpoint core would momentarily harden slightly before failing to hold a steady point source. As a visual observer, when an object loses its clean, pinpoint stellar profile like this, it becomes impossible to make a reliable, fractional magnitude estimate. Trying to estimate a value when the core is completely diffuse introduces unacceptable personal bias into the data. Therefore, whenever the pinpoint failed to re-emerge, interpolation was rejected. Instead, I estimated a flat baseline of magnitude 15.2 compared to the steady magnitude 15.0 comparison star. This rule was strictly kept throughout the campaign to guarantee that the flat floor shown in the plots represents the true, unbiased visual reality at the eyepiece.

## 5. Direct Comparisons with Space-Based Data

To verify the accuracy of my visual estimates, I compared my visual data directly with space-based observations from NASA's Transiting Exoplanet Survey Satellite (*TESS*; Ricker et al. 2015; Jenkins et al. 2016). These comparisons highlight both the strengths and the limitations of space telescopes when monitoring crowded stellar fields.

### 5.1 Pixel Blending and Spatial Resolution Limitations

The primary challenge for automated space-based photometry in this region is the presence of the close magnitude 11.7 *V* companion star. The large pixel scale of the *TESS* instrument means that light from both WR 59 and its close magnitude 11.7 *V* companion is integrated into a single photometric aperture. As documented in Section 3.2 and shown in Figure 2, this severe pixel blending results in an artificially brightened, diluted light curve that completely masks the true depth of the fading events. By contrast, physically isolating the star at the eyepiece of the 22-inch reflector at 270× magnification allows me to cleanly separate the target from the companion star. This high-magnification optical isolation provides an unblended visual record that exposes the true, deep stellar fading events that automated large-pixel pipelines fail to detect.

### 5.2 Ground-Truth Validation across Sectors 65, 100, and 102

Building on the historic Sector 38 verification fade documented in previous work (Stubbings 2026), and using early Sector 11 *TESS* data as a vital anchor during a seasonal field gap with no visual observations, this campaign focuses on direct visual overlays across Sectors 65, 100, and 102. The continuous *TESS* light curve from these sectors shows an exceptional correlation with the timing of my visual observations. It is important to note that these overlays represent a structural comparison rather than a direct match in absolute energy values. The human eye operates in the visual spectrum (~550 nm), where the fading episodes exhibit a highly pronounced visual amplitude, causing sharp, noticeable drops. In contrast, the *TESS* instrument operates in the near-infrared band (600-1000 nm), where the stellar light profile undergoes less absolute attenuation. Additionally, because the static light of the bright magnitude 11.7 *V* companion is entirely mixed into the large *TESS* pixels, the satellite's recorded variations appear heavily diluted.

### 5.3 Analysis of *TESS* Sector 65 Overlay

**Table 19. Summary of visual magnitude estimates during *TESS* Sector 65**

| Observed Event | Julian Date (JD) | Visual Mag | Interval (Days) |
|---|---|---|---|
| *TESS* Sector 65 | 2460071.8730 | 14.0 | Window Open |
| Baseline Peak | 2460075.9030 | 13.9 | 4.0300 |
| Baseline Dip | 2460077.8740 | 14.0 | 1.9710 |
| Baseline Peak | 2460078.9060 | 13.9 | 1.0320 |
| Baseline Dip / Window Close | 2460079.9290 | 14.0 | 1.0230 |

Analysis of the comprehensive visual log across this phase reveals exceptional concurrent observations between the ground-based visual data and space-based photometry. As documented in Table 19 and Figure 24, the system was executing low-amplitude, high-frequency oscillations over an 8-day window from the Tetoora Road Observatory. These persistent micro-variations were independently confirmed by Sebastian Otero's analysis of the raw *TESS* light curve data. These results are now formally included as an official reference plot directly within the International Variable Star Index (VSX) database (Watson et al. 2006). This archived space-based plot cleanly documents that short-term oscillations with rapid amplitudes reaching up to 0.1 magnitudes are intrinsically present in the *TESS* dataset, operating on an identical timeline to the visual observing runs.

The correlation between the two independent data sets is remarkably consistent. When visual estimates isolated the star at its local baseline of magnitude 14.0, the continuous *TESS* light curve flux mirrored the behaviour by settling between 0.994 and 0.995. Conversely, when I recorded minor brightenings to magnitude 13.9 (such as at JD 2460075 and JD 2460078), the *TESS* light curve registered a corresponding flux increase above 1.01. This precise monitoring demonstrates that my visual observations can successfully resolve physical micro-variations as small as 0.1 magnitudes, providing rigorous ground-truth validation for the fine structure of the light curve.

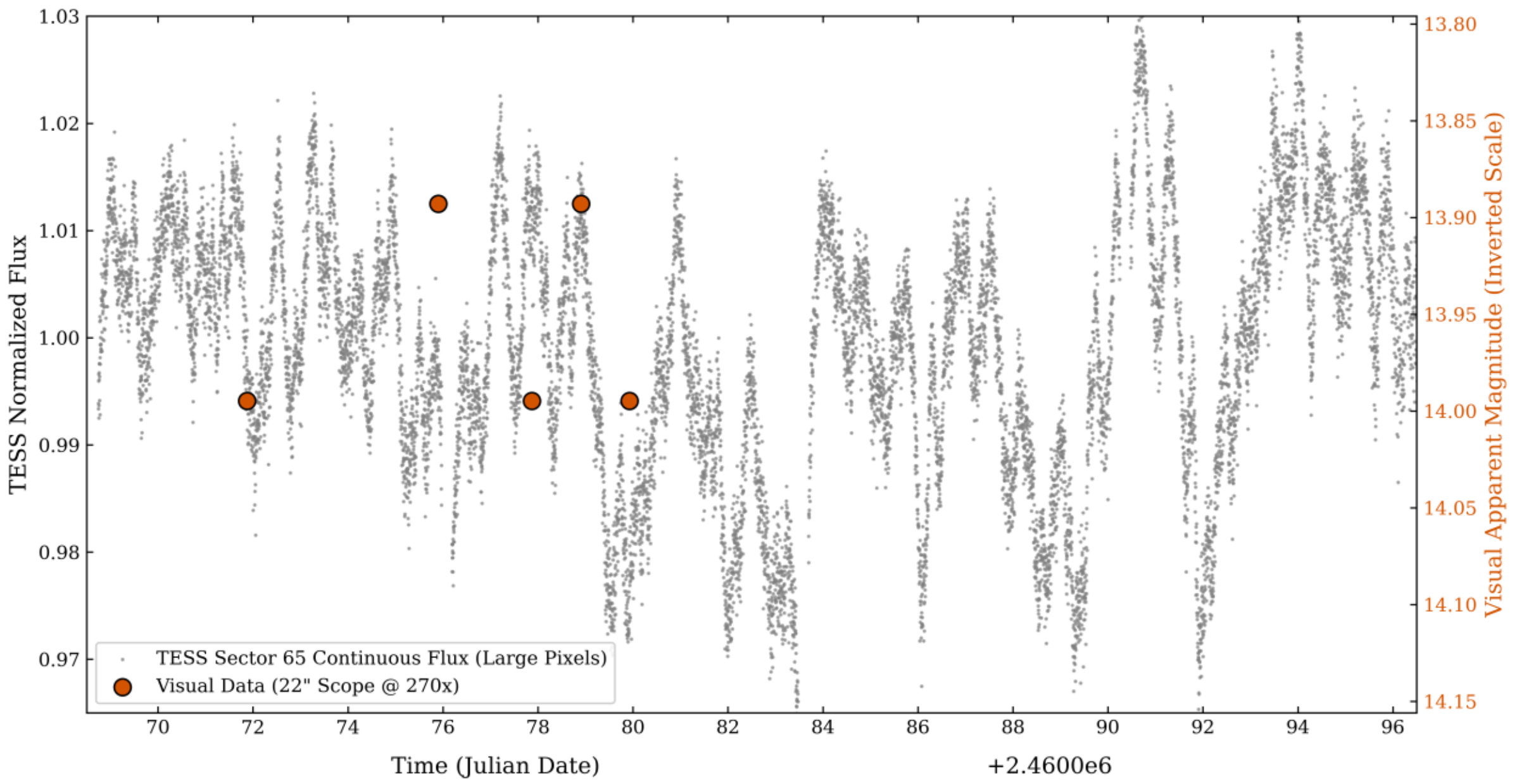

**Figure 24.** High-density unblended visual magnitude estimates (orange circles) overlaid directly onto the continuous *TESS* Sector 65 light curve (grey points) over a 27-day interval spanning JD 2460069 to JD 2460096. The visual magnitude axis on the right is inverted to match the direction of the space-based flux variations. The plot highlights remarkable visual agreement during low-amplitude oscillations, validating that the visual data can successfully resolve physical micro-variations as small as 0.1 magnitudes entirely free from the companion.

## 5.4 Analysis of *TESS* Sector 100 Overlay

**Table 20. Summary of visual magnitude estimates during *TESS* Sector 100**

| Observed Event | Julian Date (JD) | Visual Mag | Interval (Days) |
|---|---|---|---|
| *TESS* Sector 100 | 2461077.0958 | 14.1 | Window Open |
| Baseline Plateau | 2461078.0972 | 14.1 | 1.0014 |
| Baseline Plateau | 2461082.0688 | 14.1 | 3.9716 |
| Baseline Plateau | 2461084.1201 | 14.1 | 2.0513 |
| Baseline Plateau | 2461086.1167 | 14.1 | 1.9966 |
| Baseline Plateau | 2461087.0840 | 14.1 | 0.9673 |
| Baseline Plateau | 2461088.0292 | 14.1 | 0.9452 |
| Baseline Peak | 2461090.0597 | 14.0 | 2.0305 |
| Baseline Reversion / Window Close | 2461095.1222 | 14.1 | 5.0625 |

Analysis of the comprehensive visual log across this phase highlights a critical example of pixel blending and field contamination within the space-based dataset, spanning JD 2461074 to JD 2461100. Centred near JD 2461085, the continuous *TESS* light curve plot captures a severe photometric depression, with the normalised flux collapsing down to a minimum of 0.96. In an uncorrected, automated pipeline, this would be erroneously flagged as a significant fading event of the primary target. However, during high-magnification visual observations that physically isolated the star from the close magnitude 11.7 *V* companion, as documented in Table 20 and shown in Figure 25, my visual estimates remained completely flat at a slightly depressed sub-baseline shelf of magnitude 14.1 throughout this entire interval. This mismatch confirms that the deep space-based dip was an optical artefact caused by the large-pixel integration of a background fluctuation or a variation intrinsic to the companion star itself, rather than a physical plunge of the primary star.

Visual estimates recorded a brief brightening up to a baseline state of magnitude 14.0 at JD 2461090.0597. The true physical alignment returns steadily near JD 2461095.1222. After this point the *TESS* light curve continues and breaks out of its depression, and concludes near JD 2461100 with a massive upward trend towards a normalised 1.06 flux. This perfect agreement between a visual magnitude 14.1 sub-baseline and a space-based 1.00 flux level provides independent validation of the visual observer's scale. It proves that targeted visual astronomy is vital not only for detecting fades but for providing a necessary quality control filter against automated anomalies.

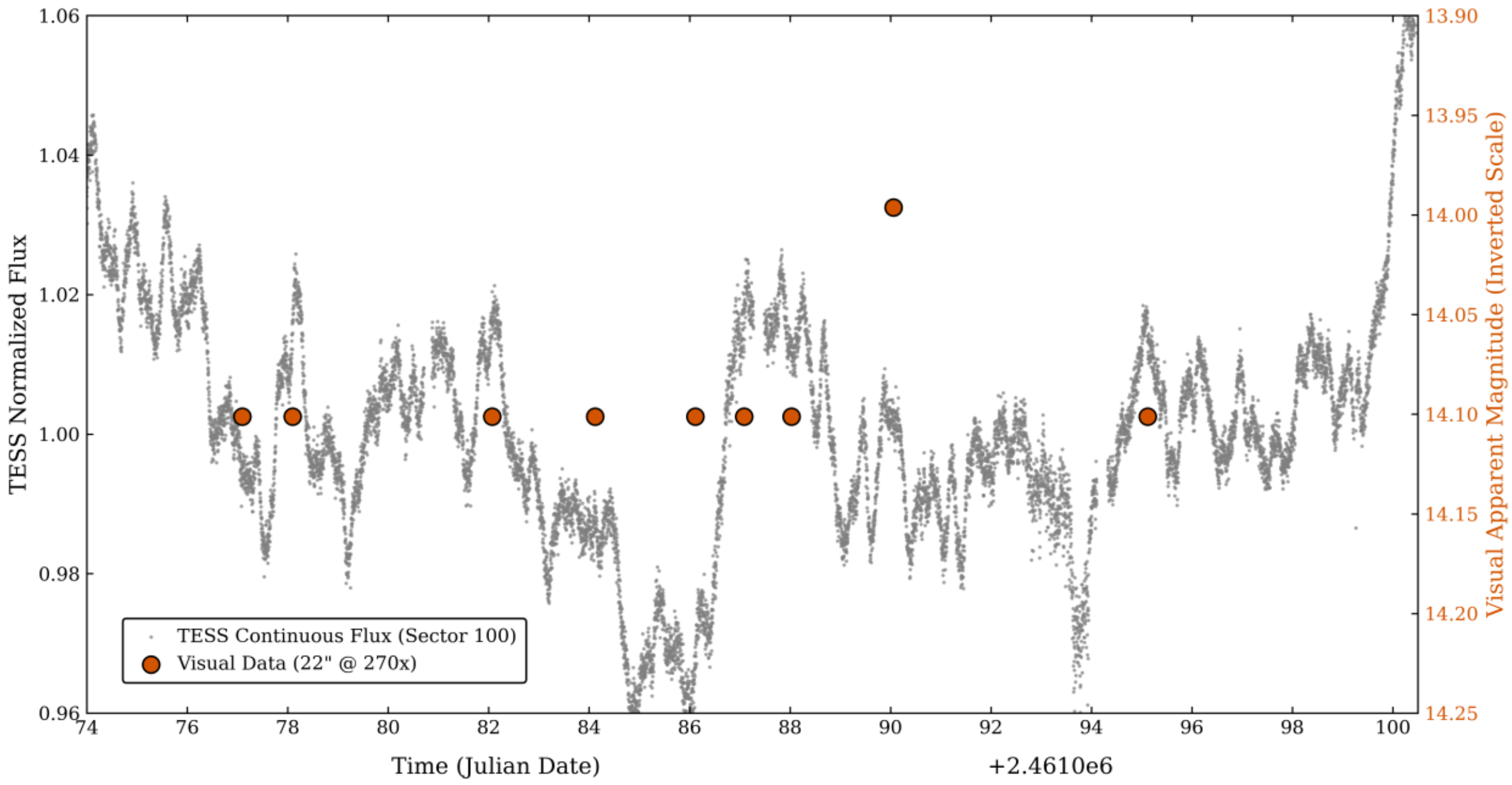

**Figure 25.** High-density unblended visual observations (orange circles) overlaid directly onto the continuous *TESS* Sector 100 light curve (grey points) across a 26-day interval. While aperture pixel blending with the close magnitude 11.7 *V* companion star creates a severe artificial space-based dip down to 0.96 near JD 2461085, the unblended visual observations remain flat at magnitude 14.1, providing an essential quality-control filter against field contamination.

### 5.5 Analysis of *TESS* Sector 102 Overlay

**Table 21. Summary of visual magnitude estimates during *TESS* Sector 102**

| Observed Event | Julian Date (JD) | Visual Mag | Interval (Days) |
|---|---|---|---|
| *TESS* Sector 102 | 2461121.0403 | 15.0 | Window Open |
| Deep Ingress Floor | 2461121.9590 | 15.0 | 0.9187 |
| Early Ascent | 2461122.0646 | 14.9 | 0.1056 |
| Recovery Rise | 2461122.9465 | 14.6 | 0.8819 |
| Recovery Hold | 2461123.0389 | 14.6 | 0.0924 |
| Upper Recovery | 2461124.0639 | 14.4 | 1.0250 |
| Approaching Baseline | 2461129.9903 | 14.1 | 5.9264 |
| At Baseline | 2461134.9215 | 14.0 | 4.9312 |
| Baseline Steady | 2461149.0785 | 14.0 | 14.1570 |
| Baseline Steady | 2461150.0590 | 14.0 | 0.9805 |
| Baseline Steady | 2461150.9160 | 14.0 | 0.8570 |
| Baseline Steady | 2461151.9347 | 14.0 | 1.0187 |
| Baseline Peak | 2461153.0910 | 13.9 | 1.1563 |
| Return to Baseline | 2461153.9215 | 14.0 | 0.8305 |
| Return to Baseline | 2461155.0639 | 14.0 | 1.1424 |
| Deep Fade Ingress | 2461156.8813 | 14.5 | 1.8174 |
| Deep Fade Hold | 2461156.9951 | 14.5 | 0.1138 |
| Deep Fade Hold | 2461156.9986 | 14.5 | 0.0035 |
| Deep Fade Ingress / Window Close | 2461157.0799 | 14.5 | 0.0813 |

Analysis of the comprehensive observational log across this phase highlights the system's structural transition immediately following Fade M, spanning JD 2461121 to JD 2461157. This interval documents the complete evolution from a deep obscured state, through an extended baseline plateau, into the initial volatile dips of the modern active phase.

The unblended visual record successfully maps the entire ingress floor and rapid egress phase of Fade M. As documented in Table 21 and Figure 26, the session opens with the star pinned flat at a minimum of magnitude 15.0 between JD 2461121.0403 and JD 2461121.9590, before executing a steep, consecutive-night egress through magnitudes 14.9, 14.6, and 14.4. This is followed by a steady near-baseline anchor at magnitude 14.1 before reclaiming its full baseline at magnitude 14.0 on JD 2461134. Upon initialising near JD 2461127, the continuous *TESS* light curve plot locks onto this recovered state, holding perfectly flat along its 1.00 normalised flux baseline matching my visual observational log records.

The perfect alignment continues through the end of the Sector 102 interval, where the visual dataset documents a brief peak at magnitude 13.9 on JD 2461153, before recording a sudden, sharp plunge down to magnitude 14.5 on JD 2461156. This close correlation provides the ultimate ground-truth validation for the observer's scale, demonstrating that targeted visual astronomy maintains complete scientific precision during long periods of quiet-state stability and multi-magnitude transitions.

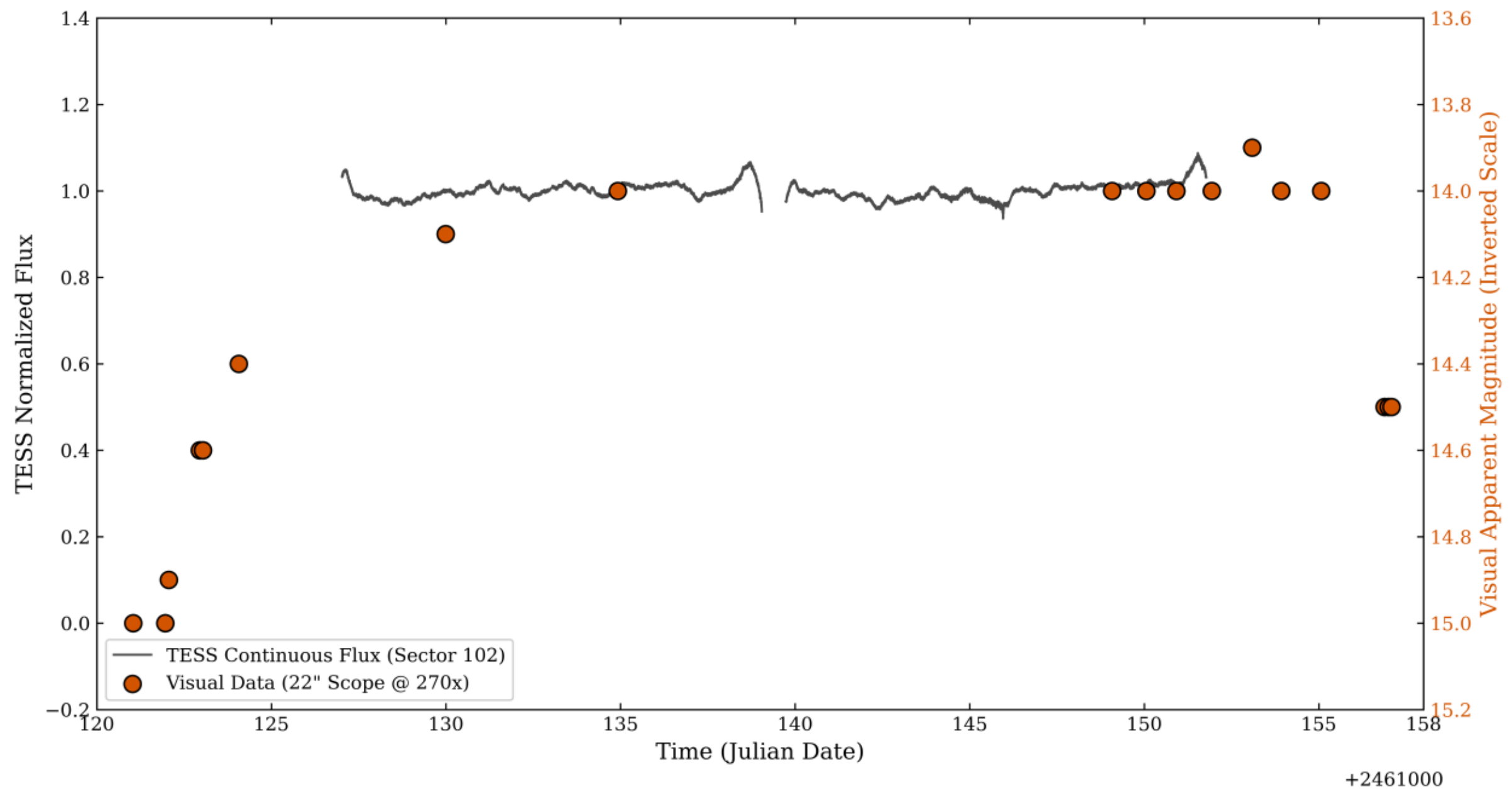

**Figure 26.** High-density unblended visual observations (orange circles) overlaid against the continuous *TESS* Sector 102 light curve (dark grey line) across a 38-day interval. The plot highlights complete visual documentation of Fade M's minimum floor at magnitude 15.0 and its subsequent egress slope where no space-based data was recorded, followed by perfect concurrent agreement across an extended, flat 1.00 normalised flux plateau, while successfully documenting the sharp structural onset of the Fade N complex.

## 6. Conclusions

The extensive visual observational campaign compiled from April 2017 to November 2026 at the Tetoora Road Observatory documents a historic behavioural transition in the WC9d Wolf-Rayet star WR 59. By maintaining continuous nightly monitoring at the eyepiece of a 22-inch telescope, this programme successfully mapped an accelerating sequence of twelve distinct fading events following the initial May 2021 discovery fade. The long-term record reveals that the system has steadily abandoned its historical, isolated multi-year fading pattern, compressing its recovery intervals down to a rapid cycle before entering the current continuous deep-fading phase. Because the automated large-pixel survey pipelines and CCD platforms remain entirely blind to this high-amplitude fading activity due to severe companion star blending, physically isolating the star at high magnification provides the only unblended dataset capable of revealing the true physical behaviour of WR 59.

The high-density visual observation blocks conducted at the absolute minimum floor provide direct, real-time evidence of highly rapid, short-timescale variations modifying the star's apparent brightness. These minute-by-minute bimodal oscillations and sub-minute micro-variations between magnitudes 15.1, 15.15, and 15.2 demonstrate the extreme variability shown in the light curve profiles. Direct ground-based overlays with NASA's *TESS* satellite photometry during quiescent phases validate the scale of my visual estimates, proving that targeted visual observations can achieve a level of precision comparable to space-based instruments while successfully bypassing field contamination.

While a definitive physical model is beyond the scope of this paper, the overall structural evolution of the master light curve suggests that the system has entered a prolonged, fundamental dust-production state. The flat, unrecovered sub-baseline plateaus and the sheer density of the current ongoing deep fading indicate that the system will likely remain heavily obscured for months to come, rather than executing a rapid, clean egress back to the baseline as witnessed during the initial 2021 episode. Because the deep magnitude 15.2 minimum of the 2026 Fade N complex remains fully ongoing, intensive visual monitoring continues to map the eventual long-term egress recovery of this unique dust-making system.

## Acknowledgements

My sincere thanks to Sebastian Otero (AAVSO/VSX) for calibrating the multi-year archival *TESS* overlay plot. I acknowledge the Mikulski Archive for Space Telescopes (MAST) for providing access to the *TESS* mission data. In addition, I gratefully acknowledge the American Association of Variable Star Observers (AAVSO) and the use of data from the AAVSO International Database contributed by observers worldwide.